\documentclass[aip, jcp,
reprint,%
]{revtex4-1}

\usepackage[utf8]{inputenc}
\usepackage[T1]{fontenc}
\usepackage{amsmath}
\usepackage{amssymb}
\usepackage{mathptmx}
\usepackage{etoolbox}
\usepackage{mathtools}
\usepackage{amsfonts}
\usepackage{wrapfig}
\usepackage{lmodern}
\usepackage{graphicx}
\usepackage{dcolumn}
\usepackage{bm}
\usepackage{url}
\usepackage[caption=false]{subfig}
\usepackage{xr}
\usepackage[hidelinks]{hyperref}

\newcommand{\RM}[1]{\MakeUppercase{\romannumeral #1{}}}
\newcommand{\fnet}{Fortnet}

\makeatletter
\def\@email#1#2{%
 \endgroup
 \patchcmd{\titleblock@produce}
  {\frontmatter@RRAPformat}
  {\frontmatter@RRAPformat{\produce@RRAP{*#1\href{mailto:#2}{#2}}}\frontmatter@RRAPformat}
  {}{}
}%
\makeatother
\begin{document}

\preprint{AIP/123-QED}

\title{\fnet{}, a software package for training Behler-Parrinello neural networks}
\author{T. van der Heide}
\affiliation{ 
Bremen Center for Computational Materials Science, University of Bremen, Bremen, Germany%
}%

\author{J. Kullgren}
\affiliation{ 
Department of Chemistry - Ångström Laboratory, Uppsala University, Box 538, 751 21 Uppsala, Sweden%
}%

\author{P. Broqvist}
\affiliation{ 
Department of Chemistry - Ångström Laboratory, Uppsala University, Box 538, 751 21 Uppsala, Sweden%
}%

\author{V. Ba\v{c}i\'{c}}
\affiliation{
Department of Physics and Earth Sciences, Jacobs University Bremen, Campus Ring 1, 28759 Bremen, Germany
}%

\author{T. Frauenheim}
\affiliation{Beijing Computational Science Research Center, 100193 Beijing, P. R. China}%
\affiliation{Shenzhen JL Computational Science and Applied Research Institute (CSAR), Shenzhen 518110, P. R. China}%
\affiliation{ 
Bremen Center for Computational Materials Science, University of Bremen, Bremen, Germany%
}%

\author{B. Aradi}
\email{aradi@uni-bremen.de}
\affiliation{ 
Bremen Center for Computational Materials Science, University of Bremen, Bremen, Germany%
}%

\date{\today}%

\begin{abstract}
A new, open source, parallel, stand-alone software package (Fortnet) has been developed, which implements Behler-Parrinello neural networks. It covers the entire workflow from feature generation to the evaluation of generated potentials, coupled with higher-level analysis such as the analytic calculation of atomic forces. The functionality of the software package is demonstrated by driving the training for the fitted correction functions of the density functional tight binding (DFTB) method, which are commonly used to compensate the inaccuracies resulting from the DFTB approximations to the Kohn-Sham Hamiltonian. The usual two-body form of those correction functions limits the transferability of the parameterizations between very different structural environments. The recently introduced DFTB+ANN approach strives to lift these limitations by combining DFTB with a near-sighted artificial neural network (ANN). After investigating various approaches, we have found the combination of DFTB with an ANN acting on-top of some baseline correction functions (delta learning) the most promising one. It allowed to introduce many-body corrections on top of two-body parametrizations, while excellent transferability to chemical environments with deviating energetics could be demonstrated. 
\end{abstract}


\maketitle

\section{Introduction}\label{sec:introduction}
Inspired by the brain of living creatures~\cite{cullochpitts} and empowered through their universal function approximation capabilities~\cite{uniapprox}, artificial neural networks (ANNs) provide an unbiased approach to a large variety of applications, were the exact mapping between input and output remains unknown or computationally unfeasible. Due to their flexible topology, optimized w.r.t.\ a particular task, ANNs have already worked their way into a wide variety of fields ranging from image processing~\cite{ann_image_processing} and speech recognition~\cite{ann_speech_recog1, ann_speech_recog2, ann_speech_recog3} to the construction of high-dimensional potential-energy surfaces~\cite{bpnn}. Neural networks have successfully expanded the field of artificial intelligence, while traditionally being prosperous in addressing intellectually demanding problems for humans defined by well-known mathematical rules, to the domain of intuitive problems whose translation into a mathematical framework of instructions remains challenging~\cite{deeplbook}. \par
Since its proposal by Walter Kohn, density functional theory (DFT)~\cite{DFT} went through an incredible success story, establishing itself as part of the standard repertoire for electronic structure calculations in the fields of chemistry, physics and materials science. However, its computationally demanding nature renders the method ill-suited for extended systems or long timescales of molecular dynamics (MD). Approximate descendants of DFT, such as density functional tight binding (DFTB)~\cite{newdftb}, strive to fill the gap between \emph{ab initio} DFT and fully empirical force-fields, but come at the expense of introducing parameters whose values need to be determined, usually by fitting to \emph{ab initio} theory or experiment. \par
Starting more than two decades ago, the first machine learning feed-forward neural network potentials were constructed to infer global properties of systems with a fixed number of degrees of freedom~\cite{ffnn_first_gen}. The major limitations of these first-generation neural network potentials included the limited number of dimensions and non-compliance with physical conservation laws regarding permutation symmetry w.r.t.\ atoms, as well as translational and rotational energy conservation. Second-generation potentials~\cite{bpnn}, which build upon the ideas of J.\,Behler and M.\,Parrinello, define a topology of atomic neural networks to construct high-dimensional potential-energy surfaces that overcome the limitations of the first generation. Due to their near-sighted nature and focus on short-range interactions, efforts have been made to extend the second-generation potentials by introducing additional sets of atomic neural networks, considering long-ranged contributions like electrostatic energies as well as non-locality~\cite{ffnn_third_gen1, ffnn_third_gen2, ffnn_fourth_gen}. \par
\emph{\fnet{}}~\cite{fortnet_zenodo} is a Behler-Parrinello neural network implementation, written in Fortran 2008. Its core features include parallelized training of fully connected feed-forward neural networks that are combined to form a Behler-Parrinello topology, the fundamental ideas of which are outlined in section~\ref{sec:bpnn}. Data driven MPI-parallelism combined with HDF5~\cite{hdf5} based I/O enables to surpass the limits of shared-memory architectures and renders \fnet{} well-suited for large datasets and network sizes. Although Fortran still plays a rather minor role in the machine learning community, overshadowed by the popularity of frameworks like Tensorflow~\cite{tensorflow}, Keras~\cite{keras}, Py-Torch~\cite{pytorch} or Scikit-learn~\cite{scikit-learn}, there are appealing characteristics inherent to a statically and strongly typed language with efficient whole-array arithmetic that support this fundamental design decision. This was also recently pointed out, and demonstrated, in the so-called neural-fortran implementation of Curcic~\cite{neural-fortran}. Since (formatted) I/O is arguably a major weakness of Fortran, a Python framework was developed to handle the generation of the HDF5 datasets as well as the extraction of all desired results out of the likewise HDF5 based output of \fnet{}. \par
With \citeauthor{ccs}~\cite{ccs} implementing the curvature constrained splines (CCS) methodology to provide a framework that enables fitting DFTB repulsive potentials with minimal human effort, another robust tool for the construction of two-body parametrizations became available. However, the tool further exposes the inability of a two-body description to capture the subtle changes in energetics when transferring between significantly different chemical environments, as obtained for silicon phases of varying coordination. To overcome these limitations, \citeauthor{ccs} managed to achieve full energetic transferability by combining self-consistent DFTB (SCC-DFTB) with a near-sighted ANN, hereafter referred to as the DFTB+ANN approach. Focus now shifts to identifying which energy terms can be trained efficiently and with sufficient accuracy. In particular, the question arises, whether the provision of an explicitly calculated DFTB energy term can improve the accuracy and the transferability of an ANN, and whether the increased computational time due to the diagonalization of the associated DFTB Hamiltonian can be justified. We have investigated this question using a dataset consisting of various periodic silicon structures, which contains, among others, challenging point defects, for which standard DFTB parametrizations fail to achieve a satisfactory agreement with PBE-DFT~\cite{DFT, gga-pbe}.

\section{Theory}\label{sec:theory}

\subsection{Feed-forward neural networks}\label{sec:ffnn}
An artificial neural network (ANN)~\cite{cieng} is a (directed) graph $G=(V,E)$, whose vertices $v\in V$ are referred to as units or neurons, connected by edges $e\in E$. The total set $V$ of vertices is divided into input $V_\mathrm{in}$, hidden $V_\mathrm{hidden}$ and output subset $V_\mathrm{out}$, that is, $V=V_\mathrm{in}\cup V_\mathrm{hidden}\cup V_\mathrm{out}$ as well as $V_\mathrm{in},V_\mathrm{out}\neq\varnothing$ and $V_\mathrm{hidden}\cap (V_\mathrm{in}\cup V_\mathrm{out})=\varnothing$. Each connection $(v,u)\in E$ is assigned a weight $w_{uv}$. Feed-forward neural networks (FFNNs) constitute an acyclic graph, i.e.\ backward connections are absent. \par
The multilayer perceptron~\cite{perceptron, cieng} is often referred to as an ordinary feed-forward neural network: Consider a graph $G=(V,E)$ satisfying $V_\mathrm{in}\cap V_\mathrm{out}=\varnothing$, $V_\mathrm{hidden}=V_\mathrm{hidden}^1\cup\dots\cup V_\mathrm{hidden}^{n-2}$ whereby $V_\mathrm{hidden}^i\cap V_\mathrm{hidden}^j=\varnothing,\, \forall\, 1\leq i<j\leq n-2$. The set of edges is given by $E\subseteq \left(V_\mathrm{in}\times V_\mathrm{hidden}^1\right)\cup\left(\bigcup_{i=1}^{n-3}V_\mathrm{hidden}^i\times V_\mathrm{hidden}^{i+1}\right)\cup\left(V_\mathrm{hidden}^{n-2}\times V_\mathrm{out}\right)$. In the case that the hidden neurons form a null set, that is $n=2,\,V_\mathrm{hidden}=\varnothing$, it holds that $E\subseteq V_\mathrm{in}\times V_\mathrm{out}$. \par
\begin{figure}[!h]
\centering
\includegraphics[width=0.90\columnwidth]{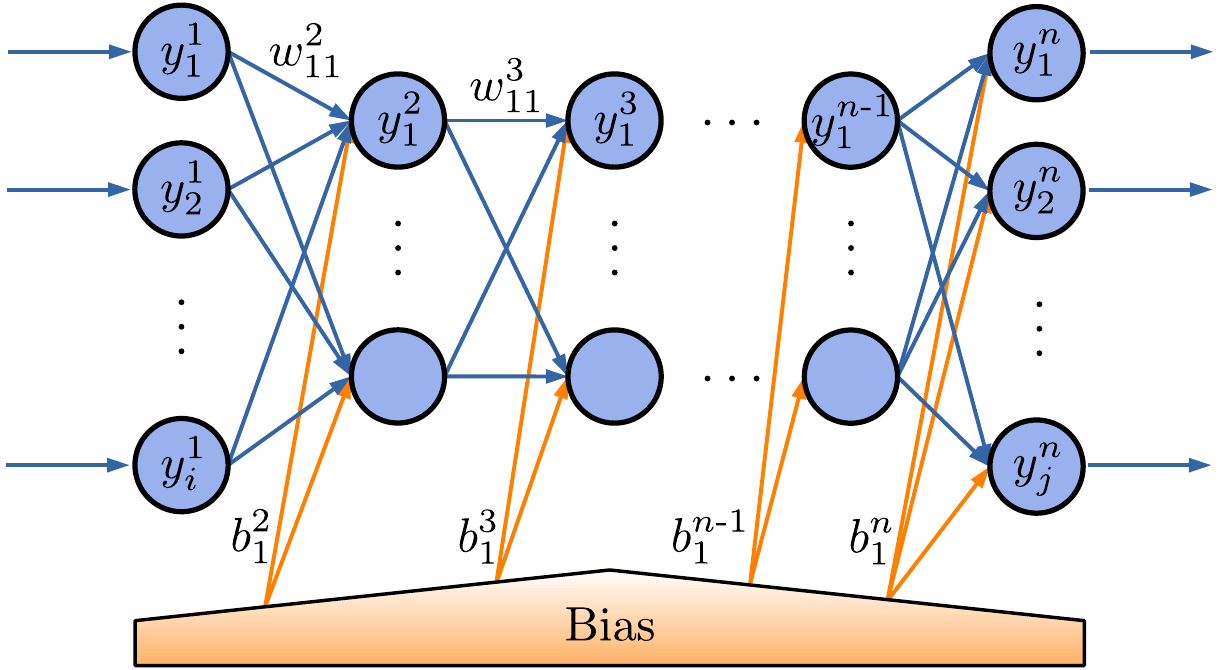}
\caption[General Multilayer Perceptron]{General $n$-layer perceptron. The $i$ input values $\bm{y}^1$ are mapped to $j$ output values $\bm{y}^n$. Between the input and output layer, $n-2\geq 0$ hidden layer, whose neurons are fully connected to the neurons of adjacent layers, further process the input by determining their weighted sum and applying a suitable activation function. Any node of layer $1<k\leq n$ is assigned an additional offset (bias) $\bm{b}^k$, shifting the argument $\bm{z}^k$ of the activation function.}
\label{fig:ml-perceptron}
\end{figure}

Therefore, the general $n$-layer perceptron maps a set of input values to a set of output values, with any additional processing taking place in $n-2\geq 0$ hidden layers. Neurons of a given layer are fully connected to the neurons of adjacent layers. Additionally, every node $u\in V_k$ of layer $1<k\leq n$ is assigned an offset (bias) parameter $b^k_u$, which shifts the argument of the respective activation function $\sigma$. Figure~\ref{fig:ml-perceptron} illustrates the general topology of such networks. \par
The actual evaluation is carried out by a forward propagation of the input through the individual layers of the network. The network input $\bm{z}^k$ of each neuron of layer $1<k\leq n$, is a weighted sum of the previous raw outputs $\bm{y}^{k-1}$ ~\cite{cieng}:
\begin{align}
\bm{z}^k = \left(\bm{w}^k\right)^\top \bm{y}^{k-1}+\bm{b}^k
\end{align}
with weight coefficients $\bm{w}^k$, biases $\bm{b}^k$, the output
\begin{align}\label{eq:fprop}
\bm{y}^k = \sigma^k\left(\bm{z}^k\right)
\end{align}
and the (usually non-linear) activation functions $\sigma^k$.\par
The quality of the predictions is usually measured by a scalar cost function $C$. The gradients of the cost function in weight-bias space can be obtained via the back-propagation algorithm. Using the error $\delta_i^k$ of neuron $i$ in layer $k$
\begin{align}
\delta_i^k \equiv \frac{\partial C}{\partial z_i^k}
\end{align}
the derivatives can be written as~\cite{bpropderivation}
\begin{align}
\bm{\delta}^k &= \nabla_{\bm{y}^k} C\odot \left(\sigma^k(\bm{z}^k)\right)' \label{eq:bp1}\\
              &= \left[\left(\bm{w}^{k+1}\right)^\top\bm{\delta^{k+1}}\right]\odot\left(\sigma^k(\bm{z}^k)\right)' \label{eq:bp2}\\
\frac{\partial C}{\partial w_{ij}^k} &= \frac{\partial C}{\partial z_i^k}\frac{\partial z_i^k}{\partial w_{ij}^k} = \delta_i^k y_j^{k-1} \label{eq:bp3}\\
\frac{\partial C}{\partial b_i^k} &= \frac{\partial C}{\partial z_i^k}\frac{\partial z_i^k}{\partial b_i^k}= \delta_i^k \label{eq:bp4}
\text,
\end{align}
where $\odot$ denotes the element-wise multiplication and prime the derivative with respect of the function argument.

The training cycle usually contains the following steps~\cite{bpropderivation}:
\begin{enumerate}
\item \textbf{Input Features:} Set the neuron activation vector $\bm{y}^1$ of the input layer, provided by the current dataset entry.
\item \textbf{Forward-Propagation:} Propagate the activation vector $\bm{y}^1$ through the subsequent layers $2\leq k\leq n$, by calculating (and caching) $\bm{z}^k$ as well as $\bm{y}^k$.
\item \textbf{Output Error:} Based on the forward-pass, determine $\bm{\delta}^n = \nabla_{\bm{y}_n} C\odot \left(\sigma^n(\bm{z}^n)\right)'$.
\item \textbf{Backward-Propagation:} Propagate the error $\bm{\delta}^k$ through the network, starting with the adjacent layer of the output layer, that is, $k=n-1,n-2,\dots ,2$.
\item \textbf{Cost Gradients:} Calculate the gradients of the cost function in the weight-bias space \\
$\curvearrowright\quad\frac{\partial C}{\partial w_{ij}^k} = \delta_i^k y_j^{k-1}$, $\frac{\partial C}{\partial b_i^k} = \delta_i^k$.
\item \textbf{Network Update:} Predict updated network parameters by using a suitable gradient-based optimizer.
\end{enumerate}

\subsection{Behler-Parrinello neural networks}\label{sec:bpnn}
\begin{figure}[!h]
\centering
\includegraphics[width=1.00\columnwidth]{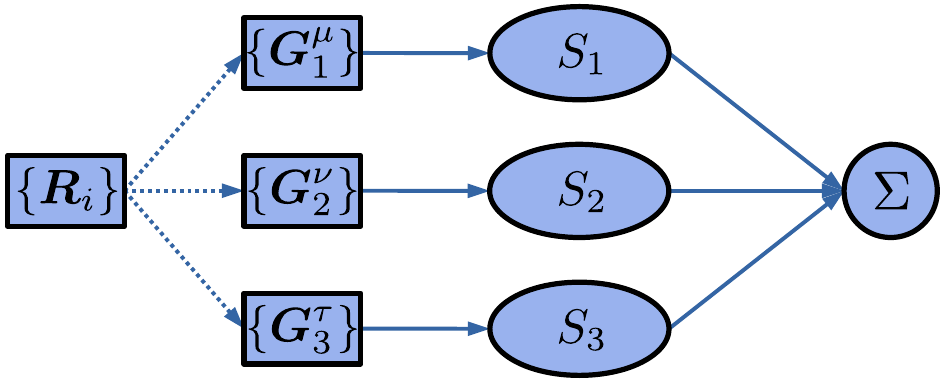}
\caption[Behler-Parrinello Neural Network]{Schematic representation of the Behler-Parrinello neural network topology~\cite{bpnn} (special case of three elements). The raw atom coordinates $\{\bm{R}_i\}$ of the dataset structures are mapped to multiple symmetry function values $\{\bm{G}_j^\mu\}$ and subsequently passed to the respective subnet $S_j$ of the corresponding element $j$. The final output $\Sigma$ follows as a summation of all atomic contributions (when fitting on system-wide properties).}\label{fig:bpnn}
\end{figure}

The neural network topology implemented in \fnet{} was proposed by J.\,Behler and M.\,Parrinello~\cite{bpnn}. It strives to overcome limitations of ordinary feed-forward neural networks (FFNNs), i.e.\ the multilayer perceptron, trained on atomic coordinates. The final result, e.g.\ total energy $E$, is decomposed into atomic contributions $E \equiv\Sigma = \sum_i E_i$ and assembled from a topology composed of multiple element related subnets $S_j$. The Behler-Parrinello neural network (BPNN), as presented by Figure \ref{fig:bpnn}, is tightly bound to the atom-centered symmetry functions (ACSF)~\cite{acsf}, which map the raw coordinates of the atoms to values that reflect their local environment w.r.t.\ neighboring atoms. By definition, the ACSF ensure that translations or rotations of a system remain energy-conserving operations, regardless of the change in the absolute coordinates, just like permuting atoms of the same element. \par
The Behler-Parrinello approach leads to a somewhat involved training process. The initial weights of each subnet are usually chosen randomly, in conjunction with vanishing bias. In preparation for the actual training iterations, the ACSF values are calculated by considering all atoms of the dataset. Subsequently, the subnets are fed one atom's ACSF set at a time, while temporarily caching the arguments of the activation functions. After feeding the network with all atoms of a certain structure, the output is obtained as a summation of atomic contributions (when fitting on system-wide properties). Once every datapoint is evaluated, the atomic gradients of a suitable cost function in weight-bias space are determined by back-propagation (cf.\ section \ref{sec:ffnn}) and condensed into a total gradient, based on which new parameters are predicted by a gradient-based optimizer.

\section{Implementation}\label{sec:implementation}

\subsection{Data parallelism}\label{sec:language}\label{sec:parallelism}

\begin{figure}[!h]
\centering
\includegraphics[width=1.00\columnwidth]{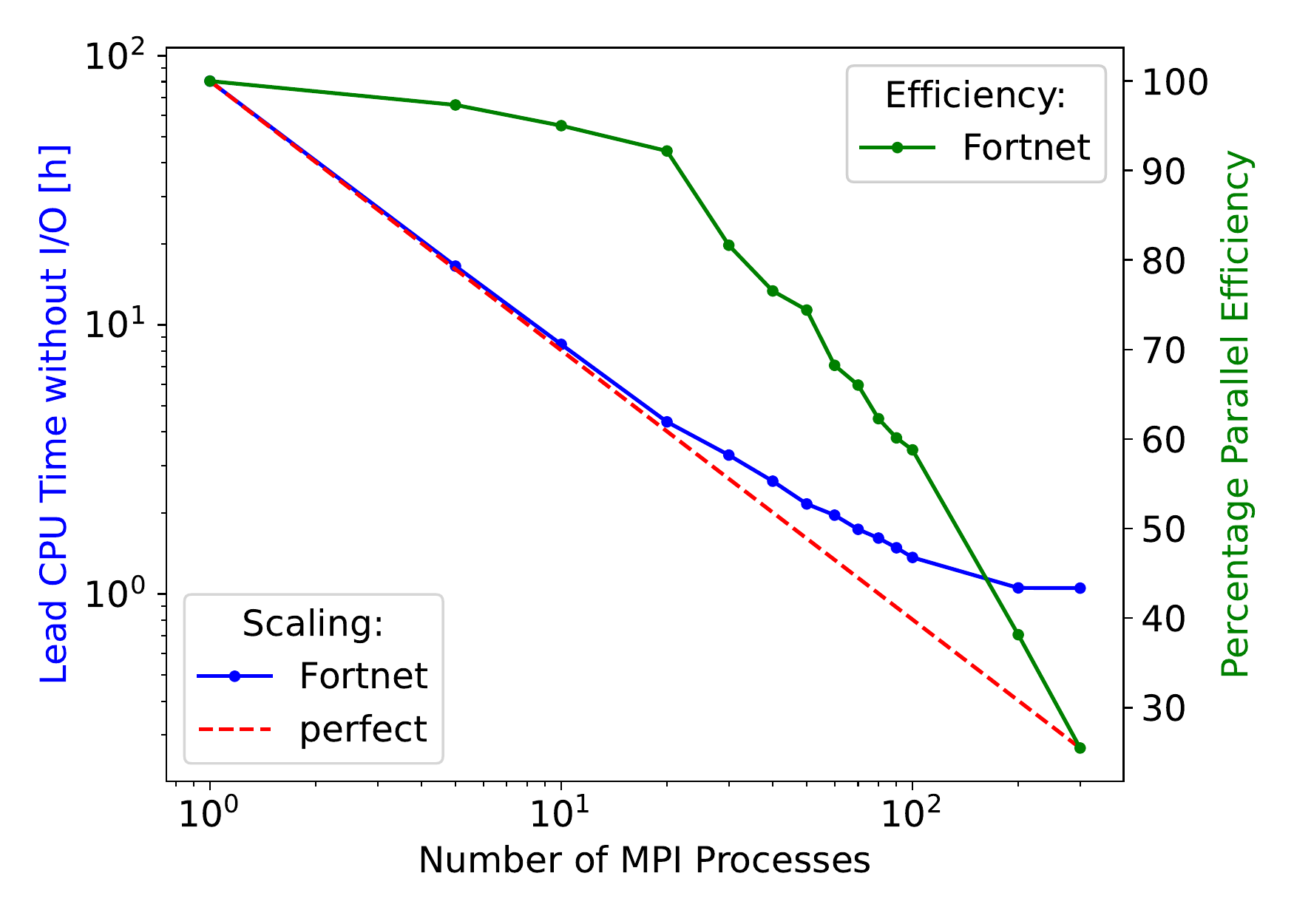}
\caption[Benchmark of \fnet{}'s MPI-Parallelization]{Illustration of \fnet{}'s MPI-parallelization. Both, the I/O-cleaned CPU time as well as parallel efficiency reduce for an increasing number of MPI processes. The computational details of the benchmark are outlined in section \ref{sec:parallelism}.}
\label{fig:benchmark}
\end{figure}

Since \fnet{} is designed to efficiently handle large datasets and network sizes, all performance critical routines are MPI-parallelized and work on both shared and distributed memory architectures. The most important parallel routines include the ACSF feature generation, network evaluation (forward-propagation) and gradient calculation (backward-propagation). \fnet{} (v0.2) was benchmarked by using a dataset of $15000$ single-element geometries, holding roughly 64 atoms per 3-dimensional periodic supercell. The sub-network architecture was chosen to be $[9\,\text{-}\,10\,\text{-}\,10\,\text{-}\,1]$, having 9 ACSF input nodes
with a cutoff of $R_\mathrm{c}=5\,\text{\AA}$, two hidden layers with 10 neurons as well as a single output cell. This topology corresponds to a total number of $221$ fitting parameters. Figure \ref{fig:benchmark} illustrates the respective I/O-less CPU times over the number of MPI processes, performing $2\cdot 10^{4}$ full-batch steepest-descent steps and initial ACSF generation. \par
Until approximately 100 processes, the parallel efficiency is quite satisfactory. Going beyond this core count, the CPU time slowly saturates, i.e.\ the serial parts of the code take over a larger relative impact on the runtime, due to the strongly accelerated gradient calculation. This leads to a reduction of the computational load and behaves as expected, considering the batch size at hand. The serial parts mainly contain the network updates by the respective optimizer. In fact, the MPI-parallelization may be \fnet{}'s greatest strength, since execution on distributed-memory machines is seamlessly possible for arbitrary core-counts.

\subsection{Program Features}\label{sec:features}
In the following, some central features of the program package are shortly presented. Further details can be found in the user documentation\footnote{see \protect\url{https://fortnet.readthedocs.io}}.

\subsubsection{Input Features}\label{sec:inputfeatures}
\paragraph{Atom-centered Symmetry Functions}\label{sec:acsf}
Atom-centered Symmetry Functions (ACSFs)~\cite{acsf} are implemented to generate input features that represent local atomic environments, by incorporating geometric information. The original paper of J.\,Behler proposed five so-called $G$-functions, parametrized by using suitable sets of $(R_c, \eta, R_\mathrm{s}, \kappa, \xi, \lambda)$ values. The function $f_\mathrm{c}$ cuts off the sphere that regards neighboring atoms in a radius of $R_\mathrm{c}$~\cite{acsf}:
\begingroup
\allowdisplaybreaks
\begin{align}\label{eq:f_cutoff}
f_\mathrm{c}(R_{ij}) =
\begin{cases}
\frac{1}{2}\left[\cos\left(\frac{\pi R_{ij}}{R_\mathrm{c}}\right) + 1\right]&\text{ for }R_{ij}\leq R_\mathrm{c} \\ 0 &\text{ for }R_{ij}> R_\mathrm{c}
\end{cases}
\end{align}
A distinction can be made between radial functions, which solely depend on interatomic distances $R_{ij}$
\begin{eqnarray}
G_i^{1,Z_1} =&& \sum_j^{|Z_1|} f_\mathrm{c}(R_{ij}) \\
G_i^{2,Z_1} =&& \sum_j^{|Z_1|} e^{-\eta (R_{ij} - R_\mathrm{s})^2} \cdot f_\mathrm{c}(R_{ij}) \\
G_i^{3,Z_1} =&& \sum_j^{|Z_1|} \cos(\kappa R_{ij}) \cdot f_\mathrm{c}(R_{ij})
\end{eqnarray}
and angular functions, that additionally incorporate a three-center defined angle~\cite{acsf} $\Theta_{ijk} = \arccos\left[(\bm{R}_{ij}\cdot \bm{R}_{ik}) / (R_{ij} \cdot R_{ik})\right]$:
\begin{eqnarray}
G_i^{4,Z_1,Z_2} =&& 2^{1-\xi} \sum_{j\neq i}^{|Z_1|}\sum_{k\neq i}^{|Z_2|} (1 + \lambda\cos\Theta_{ijk})^\xi \\
&&\times e^{-\eta (R_{ij}^2 + R_{ik}^2 + R_{jk}^2)} f_\mathrm{c}(R_{ij})f_\mathrm{c}(R_{ik})f_\mathrm{c}(R_{jk}) \nonumber\\
G_i^{5,Z_1,Z_2} =&& 2^{1-\xi} \sum_{j\neq i}^{|Z_1|}\sum_{k\neq i}^{|Z_2|} (1 + \lambda\cos\Theta_{ijk})^\xi \\
&&\times e^{-\eta (R_{ij}^2 + R_{ik}^2)} f_\mathrm{c}(R_{ij})f_\mathrm{c}(R_{ik}) \nonumber
\end{eqnarray}
\endgroup
An element-resolved scheme as presented above may result in a large number of input nodes $N_\mathrm{in}$, originating from different functions for each element (two-body ACSFs) or element-pair (three-body ACSFs) respectively:
\begin{align}\label{eq:ninputs}
N_\mathrm{in} &= N_\mathrm{elements}\cdot \sum_{i=1}^{5}N_{G^i} + \frac{N_\mathrm{elements}!}{2!\,(N_\mathrm{elements}-2)!}\cdot\sum_{i=4}^{5}N_{G^i}
\end{align}
Therefore \fnet{} provides an element-unresolved scheme whose neighbor lists are element-agnostic (cf.\ supplementary material \RM{1}, listing 1). Even considering the fact that the BPNN consists of multiple sub-networks, dedicated to the respective elements of the dataset, the raw ACSF are expected to represent contradictory data in this particular case. In order to overcome this contradiction, a similar strategy as weighted ACSF (wACSF)~\cite{wACSF} is pursued to recover discernment between different elements.
A large variety of atomic properties such as the atomic number, number of valence electrons, Hubbard $U$'s, Mulliken populations or any other suitable quantity would be conceivable to improve atomic characterization. It may be ensured that all summands have an equal sign in order to prevent cancellation of contributions. While this is not necessarily problematic, it should be handled with caution~\cite{acsf}. \par
The parametrization of the G-functions may be carried out by an automatic ACSF parameter generation scheme that aims to cover the cutoff sphere as evenly as possible, with the number of symmetry functions available. This reduces the user input to the desired cutoff radius $R_\mathrm{c}$ in conjunction with the number $N_\mathrm{rad}$ of radial $G^2$ and $N_\mathrm{ang}$ angular $G^5$ functions (cf.\ supplementary material \RM{1}, listing 2). However, each individual function can also be defined manually and for example employed to specifically extend the automatic generation scheme (cf.\ supplementary material \RM{1}, listing 3).

\paragraph{External Atomic Input Features}\label{sec:extfeatures}
Since this kind of software can never claim that all conceivable feature generators are implemented natively and to vastly improve the flexibility of \fnet{}, there is no need to solely rely on ACSF to characterize the local environment of each atom. Rather, the independent specification of external atomic input features, extracted from the dataset, is possible (cf.\ supplementary material \RM{1}, listing 4). In the context of electronic structure calculations, properties such as the atomic number, number of valence electrons, Hubbard $U$'s, Mulliken populations or any other suitable quantity may thus be included as input feature.

\subsubsection{Network Initialization Scheme}\label{sec:netinit}
The network initialization is carried out by a truncated Xavier initialization~\cite{xavierinit}. This choice primarily refers to the type of sigmoidal activation functions used to generate the results of section~\ref{sec:application}. Its main idea is to draw the weight vector $\bm{w}^k$ of layer $k$ from a normal distribution with zero mean and a variance $V$ that depends on the number of nodes $n^{k}$ of the (adjacent) layer $k$, while using vanishing bias values $\bm{b}^k$. As shown by~\citeauthor*{xavierinit}~\cite{xavierinit}, this particular choice of variance refers to the separating case between vanishing and exploding gradients and imposes variance preservation across all layers, i.e.\ $V(\bm{y}^{k}) = V(\bm{y}^{k-1})$. To obtain a suitable weight distribution, a luxury random number generator~\cite{ranlux1, ranlux2, ranlux3} draws high-quality pseudo-random numbers $x$ from an interval $x\in[x_{-}, x_{+}]$, defined by the hyperparameter $\mathcal{N}_\mathrm{min}$ of the truncation:
\begin{align*}
\mathcal{N}(\mu,\sigma^2) &= \frac{1}{\sigma\sqrt{2\pi}}e^{-\frac{1}{2}\left(\frac{x-\mu}{\sigma}\right)^2} \\
\sigma^2 \equiv V(\bm{w}^k) &= \frac{2}{n^{k-1} + n^k} \\
x_{\pm} &= \pm\sqrt{-2\sigma^2\ln\left(\mathcal{N}_\mathrm{min}\sigma\sqrt{2\pi}\right)} \\
\Longrightarrow\qquad w_{ij}^k &= g\cdot\mathcal{N}(\mu=0,\sigma^2),\quad b_i^k = 0
\end{align*}
As for $\mathcal{N}_\mathrm{min}$, the scaling factor $g$ is currently not accessible from the user perspective, although the implementation of this would be trivial, and is always chosen to be one.

\subsubsection{Forward- and Backward-Propagation}\label{sec:fbprop}
Similar to other machine learning codes, \fnet{} heavily relies on its implementation of the forward- and backward-propagation, with the network evaluation as well as the calculation of the cost-gradients in weight-bias space naturally influencing the training process. The realization of eq.\,\eqref{eq:fprop}, eq.\,\eqref{eq:bp1}, \eqref{eq:bp2}, \eqref{eq:bp3} and \eqref{eq:bp4} in terms of Fortran's array arithmetic forms the core of both algorithms. \par
The implementation was inspired by similar works~\cite{bprophist, bprop}, and the neural-fortran~\cite{neural-fortran} paper. Latter demonstrates the competitiveness of such a bare-bones implementation, compared to the Keras~\cite{keras} library with TensorFlow~\cite{tensorflow} as backend, especially highlighting the significantly reduced memory consumption. In contrast to the neural-fortran~\cite{neural-fortran} implementation, \fnet{} generalizes the back-propagation implementation to arbitrary cost-function gradients and exploits the fact that the analytical derivatives of the implemented cost functions are known. To successively optimize the weight and bias network parameters, based on the gradients obtained by back-propagation, \fnet{} provides several minimization algorithms. \par

\subsubsection{Activation Functions}\label{sec:activation}
To enable non-linear potentials, among others, logistic, arcus and hyperbolic tangent, (leaky) ReLU~\cite{relu} and softplus~\citep{softplus} transfer is available. While sigmoidal activation functions have historically had great relevance and are still frequently used for this purpose, they suffer from soft saturation and vanishing gradients, especially in inappropriately initialized~\cite{xavierinit} networks. In order to circumvent these problems, rectified linear units (ReLU)~\cite{relu} became increasingly popular, in particular due to significant speedups~\cite{relusuccess} achieved while training deep (convolutional) neural networks. The modular structure of the implementation, using abstract interfaces and procedure pointers, allows for a convenient extension of this list.

\subsubsection{Atomic and Global Training Targets}\label{sec:targets}
The implemented Behler-Parrinello topology can either be trained on atomic properties like forces and charges, or on system-wide properties such as the total energy of a geometry. A mixture of both types of targets will be subject of future developments.

\subsubsection{Loss Functions and Regularization}\label{sec:loss}
In terms of the maximum likelihood framework~\cite{deeplbook}, the minimization of the negative log-likelihood of a neural network with linear output units is carried out by minimizing the residual sum of squares, i.e.\ mean squared error. \fnet{} offers the mean squared loss as default loss function, with several alternatives:
\begin{itemize}
\item mean squared error (mse) \\
$L = \frac{1}{N}\sum_{i=1}^N \left(y_i^\mathrm{ref} - y_i^\mathrm{nn}\right)^2$
\item root mean square error (rmse) \\
$L = \sqrt{\frac{1}{N}\sum_{i=1}^N \Big(y_i^\mathrm{ref} - y_i^\mathrm{nn}\Big)^2}$
\item mean absolute error (mae) \\
$L = \frac{1}{N}\sum_{i=1}^N |y_i^\mathrm{ref} - y_i^\mathrm{nn}|$
\item mean absolute percentage error (mape) \\
$L = \frac{100}{N}\sum_{i=1}^N \Big|\frac{y_i^\mathrm{ref} - y_i^\mathrm{nn}}{y_i^\mathrm{ref}}\Big|$
\end{itemize}
Adding further loss functions is straight-forward.\par
Regularization aims to improve upon the generalization error, while leaving the training error unaltered~\cite{deeplbook}. Besides utilizing a plain early-stopping mechanism to avoid overfitting, loss-based regularization~\cite{lasso, ridge, elasticnet} puts a soft constraint onto the weights of a network, while compromising between a reduced variance and increased bias. The original cost function $C_0$ is penalized by an additional term $\tilde{C}$, whereas its influence is adjusted by a positive hyperparameter $\lambda$, that can be set from the user's perspective:
\begin{align}
C = C_0 + \lambda\tilde{C}
\end{align}
\fnet{} provides three popular loss-based regularization techniques to induce weight decay:
\begin{itemize}
\item $L^1$ - Lasso~\cite{lasso} \\
$\tilde{C} = \frac{1}{n}\sum_{i=1}^{n}|w_i|,\quad\frac{\partial\tilde{C}}{\partial w_i} = \frac{\mathrm{sgn}(w_i)}{n}$
\item $L^2$ - Ridge~\cite{ridge} \\
$\tilde{C} = \frac{1}{2n}\sum_{i=1}^{n}w_i^2,\quad\frac{\partial\tilde{C}}{\partial w_i} = \frac{w_i}{n}$
\item Elastic Net~\cite{elasticnet} \\
$\tilde{C} = \frac{1}{n}\left(\frac{1-\alpha}{2}\cdot\sum_{i=1}^{n}w_i^2 + \alpha\cdot\sum_{i=1}^{n}|w_i|\right) \\
\frac{\partial\tilde{C}}{\partial w_i} = \frac{1}{n}\left[(1-\alpha)\cdot w_i + \alpha\cdot\mathrm{sgn}(w_i)\right]$
\end{itemize}
$L^1$ (lasso) an $L^2$ (ridge) regularization encourage continuously shrinking weight coefficients, while training the network. The qualitative difference of lasso regression w.r.t.\ ridge regression is the promotion of sparsity, i.e.\ automatic feature selection~\cite{elasticnet}, whereas an $L^2$ penalty is not capable of deactivating neurons. Notwithstanding this, there are cases in which the feature selection by lasso empirically proved unreliable~\cite{elasticnet} and the prediction accuracy of ridge-penalized networks prevails. To combine the strengths of both approaches, the elastic net~\cite{elasticnet} regularization got introduced as a mixture of lasso and ridge regression. The special cases $\alpha = 0$ and $\alpha = 1$ are equivalent to $L^2$ and $L^1$ regularization respectively.

\subsubsection{Atomic Forces}\label{sec:forces}
Atomic forces, i.e.\ the negative gradients of the total energy $E$ w.r.t.\ the coordinates $R_{l}^\alpha,\alpha = (x,y,z)$ of atom $l$, may be determined numerically or analytically. Former case is based on the central finite difference:
\begin{align}
F_{l}^\alpha = -\frac{\partial E}{\partial R_{l}^\alpha} \approx \frac{E(R_{l}^\alpha - \Delta R_{l}^\alpha) - E(R_{l}^\alpha + \Delta R_{l}^\alpha)} {2\Delta R_{l}^\alpha}
\end{align}
Although the associated computation of $6N$ ACSF sets per datapoint of $N$ atoms is MPI-parallelized, an analytical computation based on the BPNN's forward derivative~\cite{forwardderiv} is more advantageous. Since the ACSF are explicitly known and differentiable, the analytical force components emerge by applying the chain rule~\cite{acsf}:
\begin{align}
F_{l}^\alpha = -\sum_{i=1}^N \frac{\partial E_i}{\partial R_{l}^\alpha} = -\sum_{i=1}^N \sum_{j=1}^{M_i} \frac{\partial E_i}{\partial G_{i,j}}\frac{\partial G_{i,j}}{\partial R_{l}^\alpha}
\end{align}
The second sum treats $M_i$ ACSF functions that represent the local environment of the $i$th atom. The first derivatives~\cite{forwardderiv} of the network output $\bm{y}^k(\bm{x}) \in \mathbb{R}^{m_k}$ w.r.t.\ the input features $\bm{x} \in \mathbb{R}^{m_1}$ may be defined as Jacobian matrix $J^n(\bm{x}) \in \mathbb{R}^{(m_n\times m_1)}$ of layer $k = n$:
\begin{align}
J^n(\bm{x}) &= \vec{\nabla}\bm{y}^n(\bm{x}) \notag\\
&= \mathrm{diag}\left[\sigma'^n\left(\left(\bm{w}^n\right)^\top \bm{y}^{n-1}(\bm{x}) + \bm{b}^n\right)\right] \notag\\
&\phantom{\ \mathrm{diag}}\times \left[\left(\bm{w}^n\right)^\top J^{n-1}(\bm{x})\right]
\end{align}
To obtain $J^{n-1}(\bm{x})$, a recursion is carried out $\forall k = 2,\dots n-1$, originating from the identity $J^1(\bm{x}) = I\in \mathbb{R}^{m_1\times m_1}$. Explicit expressions for the ACSF derivatives can be found in appendix~\ref{sec:appendix-acsf-prime}.

\begin{figure}[!h]
\centering
\includegraphics[width=1.00\columnwidth]{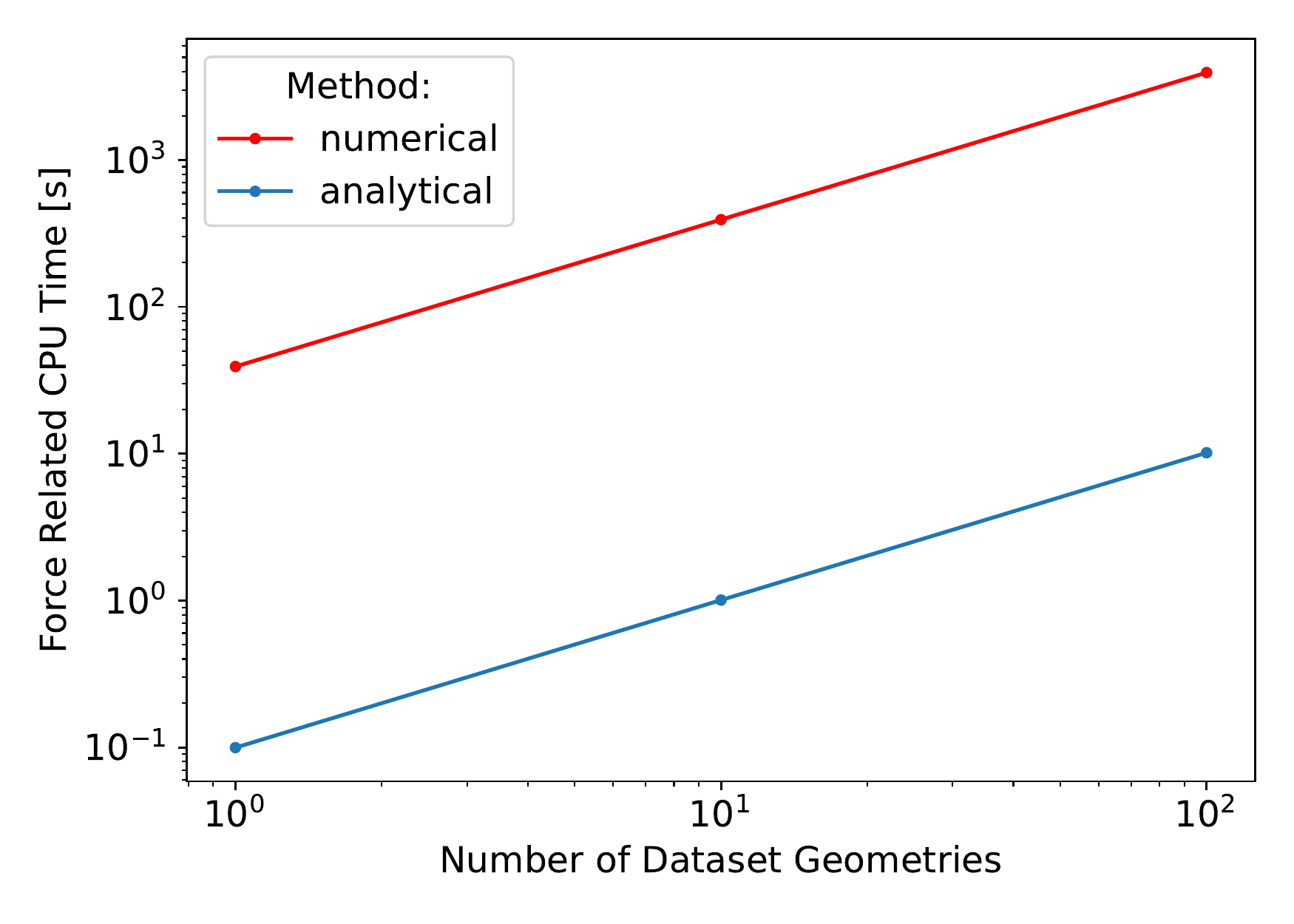}
\caption[Benchmark of \fnet{}'s Force Analysis Methods]{Comparison between \fnet{}'s analytical and numerical force analysis, in terms of force related CPU time.}
\label{fig:acsf-prime}
\end{figure}

We have benchmarked the analytical force calculation by using datasets of $10^{0}$, $10^{1}$ and $10^{2}$ silicon carbide geometries, holding 64 atoms per 3-dimensional periodic supercell. The sub-network architectures were chosen to be $[32\,\text{-}\,3\,\text{-}\,3\,\text{-}\,1]$, having 32 ACSF input nodes with a cutoff of $R_\mathrm{c}=5\,\text{\AA}$, two hidden layers with 3 neurons as well as a single output cell. This topology corresponds to a total number of $230$ fitting parameters. Figure \ref{fig:acsf-prime} illustrates the relevant CPU times over the number of dataset geometries, performing analytical and numerical force analysis. \par
The analytical and numerical force related, I/O-cleaned CPU time increases linearly with the number of dataset geometries. However, the total timings of the two methods differ significantly, showing the vast advantage of analytical expressions for the Jacobi matrix of the network and the ACSF derivatives w.r.t.\ the atomic coordinates, over numerical central finite differences.

\subsubsection{Generating a Dataset}\label{sec:fnetdata}
As already mentioned in section~\ref{sec:introduction}, a Python framework handles the composition of compatible datasets and enables the extraction of selected results after a \fnet{} run has been completed. Thereby it relies on the strength of the Python package Atomic Simulation Environment (ASE)~\cite{ase} to represent geometric information that may have previously been extracted from the outputs of the employed simulation package. The actual dataset generation is as simple as illustrated by code listing $5$ of the supplementary materials. By invoking the referenced code, an HDF5 dataset will be written to disk, that includes geometric information for later ACSF generation, external atomic input features as well as system-wide training targets.

\subsubsection{Extracting Results}\label{sec:fnetout}
Analogous to the dataset generation, extracting results from the HDF5 output is carried out via a Python layer (cf.\ supplementary material \RM{1}, listing 6). To fetch desired entries the class provides several properties that may be extracted, including but not limited to the mode of the run that produced the output file (validation or pure prediction), the number of datapoints the network was trained on, the type of training targets (atomic or system-wide), the predictions of the network potential as well as corresponding targets if provided.

\subsubsection{Interfacing with external codes}
\fnet{} can currently be interfaced with other software packages using foremost file communication. (Future developments will focus on establishing an API that enables library interfacing to reduce the I/O overhead.) The external driver creates necessary input files for \fnet{} and starts an individual \fnet{} program for each of the inputs. After \fnet{} has finished, the driver analyses the created output files and extracts the necessary information from those. A file communication based interface to the Atomic Simulation Environment (ASE) package\cite{ase} is available.

\section{DFTB Repulsive Parametrization}\label{sec:application}

In this section we demonstrate the capabilities of the \fnet{} package by training and evaluating ANNs for a combined DFTB+ANN approach, where the force-field like contribution of the DFTB method is replaced or corrected by an ANN.

\subsection{DFTB basics}\label{sec:dftb}
The density functional tight binding (DFTB)~\cite{newdftb} fills the gap between \emph{ab initio} DFT electronic structure methods and fully empirical force-fields in the domain of materials and chemical modeling. Foulkes and Haydock~\cite{secondordercorr} represented the ground-state density $\rho(\bm{r})$ as a perturbed reference density $\rho_0(\bm{r})$, i.e.\ emerging from a superposition of neutral atomic densities. The self-consistent charge (SCC) extension~\cite{sccextension} originates from a second-order expansion of the Kohn-Sham total energy w.r.t.\ atomic charge fluctuations $\delta \rho$ around a reference density $\rho_0$:
\begin{align}
E[\rho_0+\delta \rho] &=& E^0[\rho_0] + E^1[\rho_0,\delta \rho] + E^2[\rho_0,\delta \rho^2]\\
&=& E_\mathrm{rep}[\rho_0] + E_\mathrm{elec}[\rho_0, \delta \rho, \delta \rho^2]  
\text.
\end{align}
The first and second order (so called electronic) terms are calculated  by expanding the Kohn-Sham states into a linear combination of atomic orbitals (LCAO) using a small valence-only basis set \cite{dftbbeginner}. The resulting Hamiltonian is treated in a two center approximation, allowing for a very efficient construction using distance dependent pretabulated Hamiltonian and overlap integrals. By solving the resulting generalized eigenvalue problem, one obtains eigenvectors as well as eigenvalues. Since SCC-DFTB considers charge fluctuations that depend on the eigenvectors, a self-consistent treatment is necessary to obtain them.\par
The zeroth order (so called repulsive) term, is usually represented as a sum of atomic pair-potentials, in line with the two-center approximation in the Hamiltonian:
\begin{align}
E_\mathrm{rep}(\rho_0, \{\bm{R}_{AB}\}) = \frac{1}{2}\sum_{A}\sum_{B\neq A}V_{AB}^\mathrm{rep}\left(|\bm{R}_A - \bm{R}_B|\right)
\end{align}
The atomic pair-potentials are fitted using ab initio or experimental reference data, in order to compensate the errors resulting from the approximate treatment of the electronic terms. This way, the DFTB method offers efficient, atomistic, quantum mechanical simulations with reasonable accuracy.

\subsection{Limitations}\label{sec:curstate}
The restrictions imposed by the two-body representation of the repulsive energy lead in certain cases to transferability issues regarding energetics in very different chemical environments. Prominent examples include the borg-0-1~\cite{borgset} parametrization with an inaccurate 2D-3D transition of boron clusters~\cite{boronfail}, the incorrect stability order of the $\mathrm{ZnO}$ wurtzite and $\mathrm{NaCl}$ phases~\cite{znofail} provided by the znorg-0-1~\cite{znorgset} parameters or the silicon polymorph issue, as already observed by~\citeauthor{siliconfail}~\cite{siliconfail}. To some extent, these mis-descriptions can be cured, for example by fitting to a dataset containing different polymorphs. In the ZnO case this led to a corrected stability order~\cite{znofail} but always compromises between transferability and accuracy, resulting from the inevitable limited geometric information and flexibility of the functional foundation. \par
Recently, \citeauthor{ccs}~\cite{ccs} implemented the curvature constrained splines (CCS) framework and demonstrated that full energetic transferability between the silicon phases may be achieved by combining self-consistent DFTB with a near-sighted artificial neural network.

\subsection{Training on Total Energy}
\subsubsection{Objective}\label{sec:objective}
Our goal is to study various different training and evaluation strategies in order to obtain DFT total energies from an ANN. We investigate and compare following scenarios:
\begin{enumerate}
\item Training on $E_\mathrm{tot}^\mathrm{DFT}$, the DFT total energies, enabling the ANN to predict the total energies directly.
\item Training on $E_\mathrm{tot}^\mathrm{DFT} - E_\mathrm{elec}^\mathrm{DFTB(pbc)}$, so that the ANN predicts repulsive energies on top of the electronic energies obtained from the pbc-0-3~\cite{pbc} set. This set is known for having a good overall performance in the energetic description of silicon.
\item Training on $E_\mathrm{tot}^\mathrm{DFT} - E_\mathrm{elec}^\mathrm{DFTB(siband)}$, so that the ANN predicts repulsive energies on top of the electronic energies obtained from the siband-1-1~\cite{siband1} parameter set. This set provides a very accurate description of the silicion band structure.
\item Training on $E_\mathrm{tot}^\mathrm{DFT} - E_\mathrm{tot}^\mathrm{DFTB(pbc)}$, so that the ANN becomes capable to compensate the errors in the total energies obtained from the pbc-set.
Compared to scenario 2, here the repulsive energy of the pbc-set is taken into account and serves as a kind of baseline.
\item Training on $E_\mathrm{tot}^\mathrm{DFT} - E_\mathrm{tot}^\mathrm{DFTB(siband+CCS)}$, where a CCS type \cite{ccs} repulsive potential fitted for the siband set serves as a baseline. (The siband set originally does not contain repulsive energy contributions.) We have fitted the repulsive energy by incorporating the entire set of training systems.
\end{enumerate}
By covering the quantum mechanically relevant part, including long-range electrostatics, through the DFTB-Hamiltonian in approaches 2--5, the ANN is left to predict a near-sighted classical potential. With a fully quantum mechanical treatment of the electrons (at least those being designated to the valence space), we could expect to find an improved transferability to other system sizes, new chemical environments or electronic temperatures etc. \par
Our investigation primarily aims at addressing whether the time penalty caused by the diagonalizations of the DFTB Hamiltonian is justified by a substantially improved transferability. The aim is to compile a dataset that allows to investigate the overall accuracy of networks trained on the various energy terms presented above. The calculations examine silicon in the bulk phase as a model system. Special attention has been paid to the description of intrinsic point defects such as the vacancy or self-interstitial configurations, which are not exactly well described in the domain of the DFTB method and thus offer a great potential for improvement. The reference will be GGA-PBE~\cite{gga-pbe} DFT as implemented by the Vienna Ab initio Simulation Package~\cite{VASP4}, trying to reproduce the total energy and formation energy of several point defects. \par
In particular, once the trained networks are available, it is necessary to investigate which ones exhibit the lowest generalization error. Here, the transferability to larger supercells roughly holding eight times the atom count shall be tested. A satisfactory transferability to larger systems is of central importance, since the dataset generation would be more efficient if only moderately sized systems need to be included.

\subsubsection{Dataset}\label{sec:dft-dftb-dataset}
The dataset holds a total number of $6306$ unique homonuclear geometries with $64\pm 1$ atoms per 3-dimensional periodic silicon supercell. The dataset is divided into three major categories, serving different purposes during training:
\begin{enumerate}
\item relaxed point defects
\item low temperature molecular dynamics
\item randomly rattled point defects
\end{enumerate}
The first subset regards relaxed point defects, such as a vacancy, $\langle 100 \rangle$-split interstitial, $\langle 110 \rangle$-split interstitial, hexagonal-site interstitial or tetrahedral-site interstitial. Other defects, which should be predicted later, had been omitted from the training set to test the generalization capability of the neural network. To ensure that the included structures are assigned enough impact onto the training, the corresponding gradients during backpropagation were homogeneously weighted by a factor of 200. \par
The second subset contains geometries of low temperature $T=500\,\mathrm{K}$ bulk silicon molecular dynamic (MD) simulations. Every 50th step, the resulting geometry has been extracted, on which basis a single-point calculation was carried out to obtain the total energy terms. The resulting geometries generate a dense sampling of the nearly perfect silicon supercells, with minor deviations mostly following the normal modes. \par
The third subset contains randomly distorted (rattled) geometries of the relaxed point defects. Their generation utilized normal distributed random numbers to rattle the coordinates of those atoms that are directly involved in building the defect, whereby surrounding atoms were masked to remain unaltered. This certainly results in structures far away from the energetic minimum of the relaxed geometries, which actually is intentional so that the network has ``seen'' a wide range of fingerprints and energies during the training.
\begin{figure}[!h]
\includegraphics[width=1.00\columnwidth]{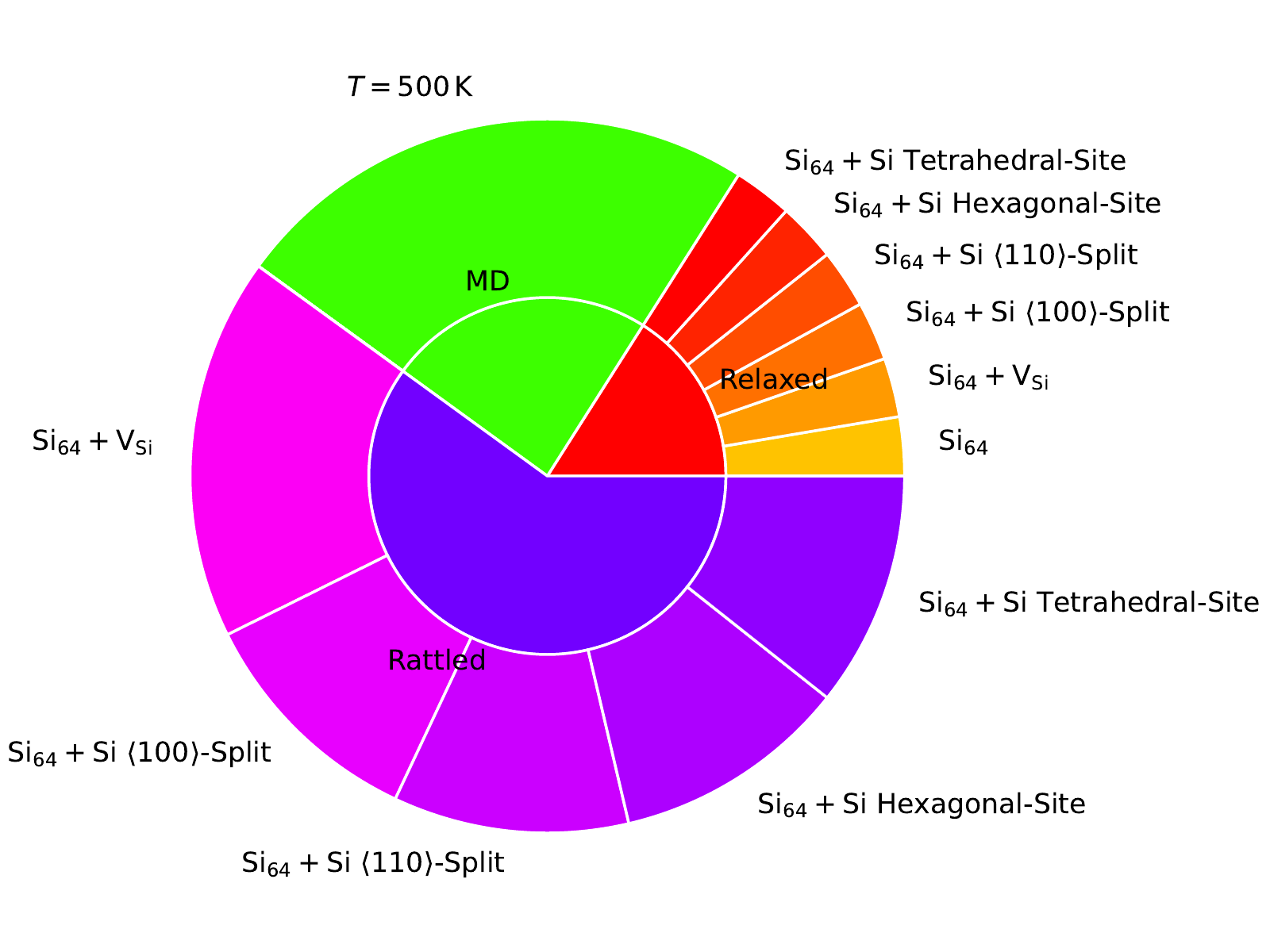}
\caption[Silicon Dataset Components]{Contributions to the training dataset deployed to investigate the achieved accuracy when fitting a BPNN to various system-wide energy terms. The set holds a total number of $6306$ unique geometries ($7500$ if the individual weighting of certain structures is considered) with $64\pm 1$ atoms per 3-dimensional periodic supercell.}
\label{fig:silicon-defect-dataset}
\end{figure}
For the silicon vacancy, a higher count of geometries had been included, as it was found that the network tend to have some difficulties learning the energetics of its four interacting dangling bonds. The described compilation of the training dataset is reflected by Figure~\ref{fig:silicon-defect-dataset}. \par
Every single-point calculation was performed by GGA-PBE DFT as well as DFTB with a $\Gamma$-centered $4\times 4\times 4$ Monkhorst-Pack~\cite{MonkhorstPack} k-point sampling for the 64 atom cells, and with a $2\times 2\times 2$ sampling for the 512 atom supercells in the test set. The DFT simulations were carried out by the Vienna Ab initio Simulation Package~\cite{VASP1, VASP2, VASP3, VASP4} with the GGA-PBE~\cite{gga-pbe} exchange-correlation functional, plane-wave basis cutoff of $500\,\mathrm{eV}$, self-consistent field tolerance of $10^{-6}\,\mathrm{eV}$ and four explicit valence electrons. In analogy, the SCC-DFTB calculations were performed by the DFTB+~\cite{newdftb} package, utilizing two established parametrizations: pbc-0-3~\cite{pbc} featuring a minimal s,p-basis and siband~\cite{siband1} with an extended basis that further includes d-orbitals.

\subsubsection{Transferability of DFTB-Hamiltonian aided DFT Total Energy Reproduction}\label{sec:dft-targets-training}
In this section we compare different possibilities to provide energetic corrections to the DFTB Hamiltonian with DFT total energies serving as reference. A BPNN with a [9-3-3-1] silicon sub-network and $N_\mathrm{rad}=5$ radial $G^2$ functions, $N_\mathrm{ang}=4$ angular $G^5$ functions and a cutoff of $R_\mathrm{c}=5\,\text{\AA}$, parametrized according to \fnet{}'s automatic parameter generation scheme, was chosen to carry out the investigations. A logistic, sigmoidal transfer function was deployed to calculate the neuron activations. In an early-stopping manner, the termination criterion of the training runs was provided by the arrival at the minimum prediction loss, whereas the runs were initially performed until full convergence. Having scanned the relevant iteration space, the network state of the corresponding early-stopping iteration got extracted.

\begin{figure}[!h]
\centering
\includegraphics[width=1.00\columnwidth]{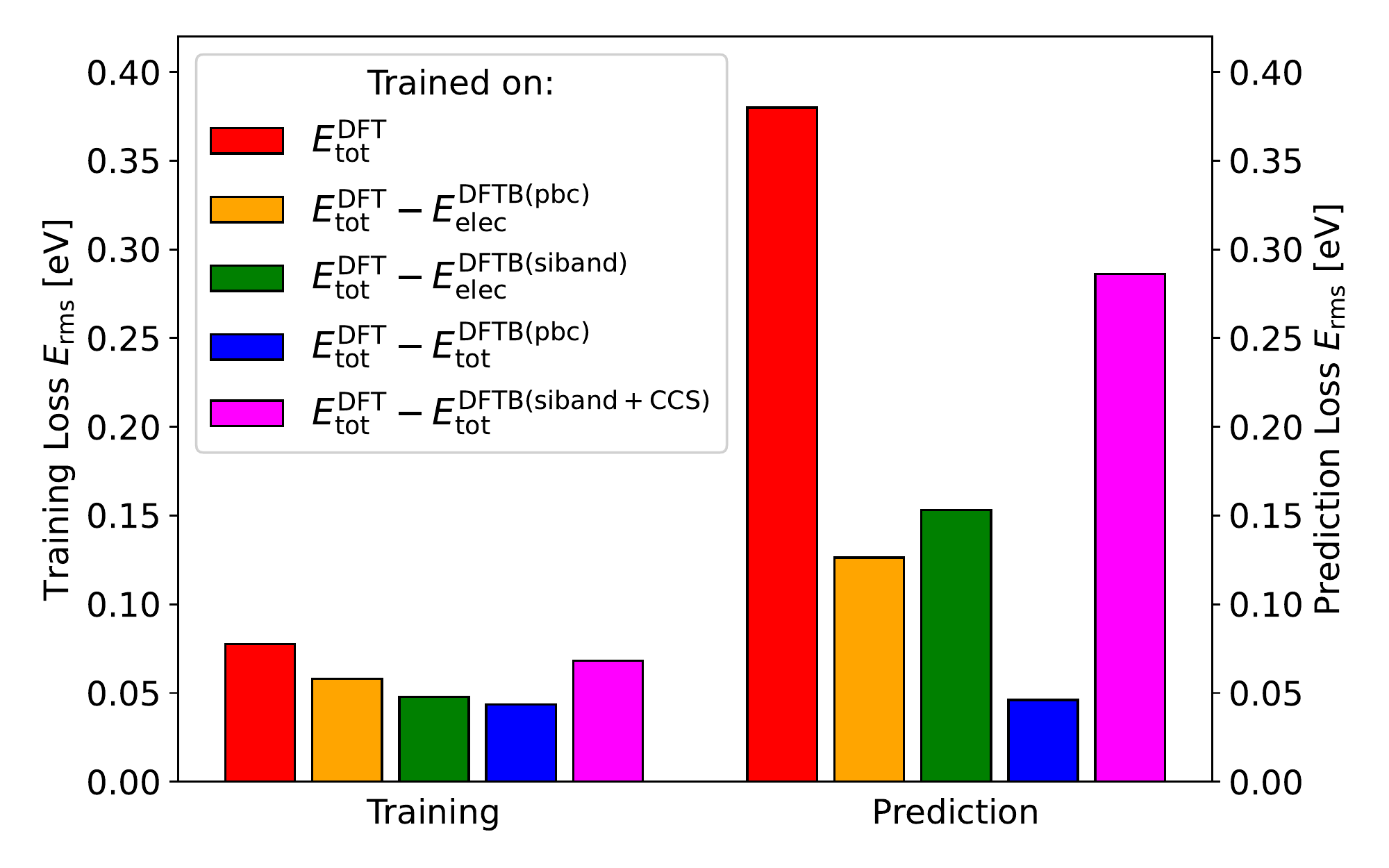}
\caption[Training and Prediction Loss when Fitting to Various Energy Targets that Reproduce DFT]{Training and prediction loss when fitting a BPNN of fixed topology to various energy targets with DFT reference. The bars correpsond to the objectives described in Section \ref{sec:objective}.}
\label{fig:silicon-dftb-training-rms-scale}
\end{figure}

Our study focuses on the transferability of the trained networks. As described above, our training set is composed of various 64 atom Si supercells. The test set, on the other hand, consists of 512 atom silicon supercells only: the perfect supercell and five supercells with various point defects. Figure \ref{fig:silicon-dftb-training-rms-scale} displays the resulting root mean square training and prediction loss values.

As can be seen in Figure \ref{fig:silicon-dftb-training-rms-scale}, the network trained directly on the total DFT energy performs worse in the training, and also  shows the worst transferability when evaluated on the larger supercells. During training, the network had to assign the structural information to the energetics, which for our training dataset is largely determined by the long-ranged defect-defect interactions. The energetic situation is qualitatively different in the larger supercells with much longer defect-defect distances, which explains the vast discrepancy between training and prediction loss.

The prediction capability of the ANN improves considerably, if the DFTB Hamiltonian is used to describe the electronic properties of the system and the ANN only provides a correction on top of it. We have followed two different strategies, by either replacing the entire repulsive contribution in DFTB with the ANN, or by using a simple ``baseline'' two-body repulsive potential in DFTB and correcting the resulting DFTB total energies with the ANN. As described above, both strategies has been tested with two different parametrizations, one with a good overall perfomance (pbc set) and one which was tuned to deliver a very precise band structure (siband set). The yellow and green bars in Figure \ref{fig:silicon-dftb-training-rms-scale} show the perfomance of the two sets, when the ANN replaces the entire repulsive contribution. With a prediction loss that is only roughly a factor of two higher than the training loss, these results are already satisfactory, especially considering that the prediction supercells contain eight times more atoms when compared to the training structures. If we include a two-body baseline repulsive potential (pink and blue bars in Figure \ref{fig:silicon-dftb-training-rms-scale}), the results improve even further, if the the repulsive contribution is derived for a well balanced electronic parametrization, as in the case of the pbc set (blue bar in Figure \ref{fig:silicon-dftb-training-rms-scale}). In this case the network solely provides minor corrections that are impossible to resolve with the limited functional form of two-center repulsive potentials. As these corrections are short-ranged, it results in a superior transferability, where the performance of the ANN, which was trained on $64\pm1$ atom supercells, is almost unchanged when applied to $512\pm1$ atom supercells. To us, this seems to be the most promising DFTB based machine learning workflow. However, the success of this strategy apparently needs a well balanced parametrization set. When applying it to the siband set, which was highly tuned (over-optimized) to deliver very accurate valence and conduction band edge states, the transferability is far more worse than with the balanced pbc parametrization (compare pink and blue bars in Figure \ref{fig:silicon-dftb-training-rms-scale}).

\begin{figure}[!h]
\centering
\includegraphics[width=1.00\columnwidth]{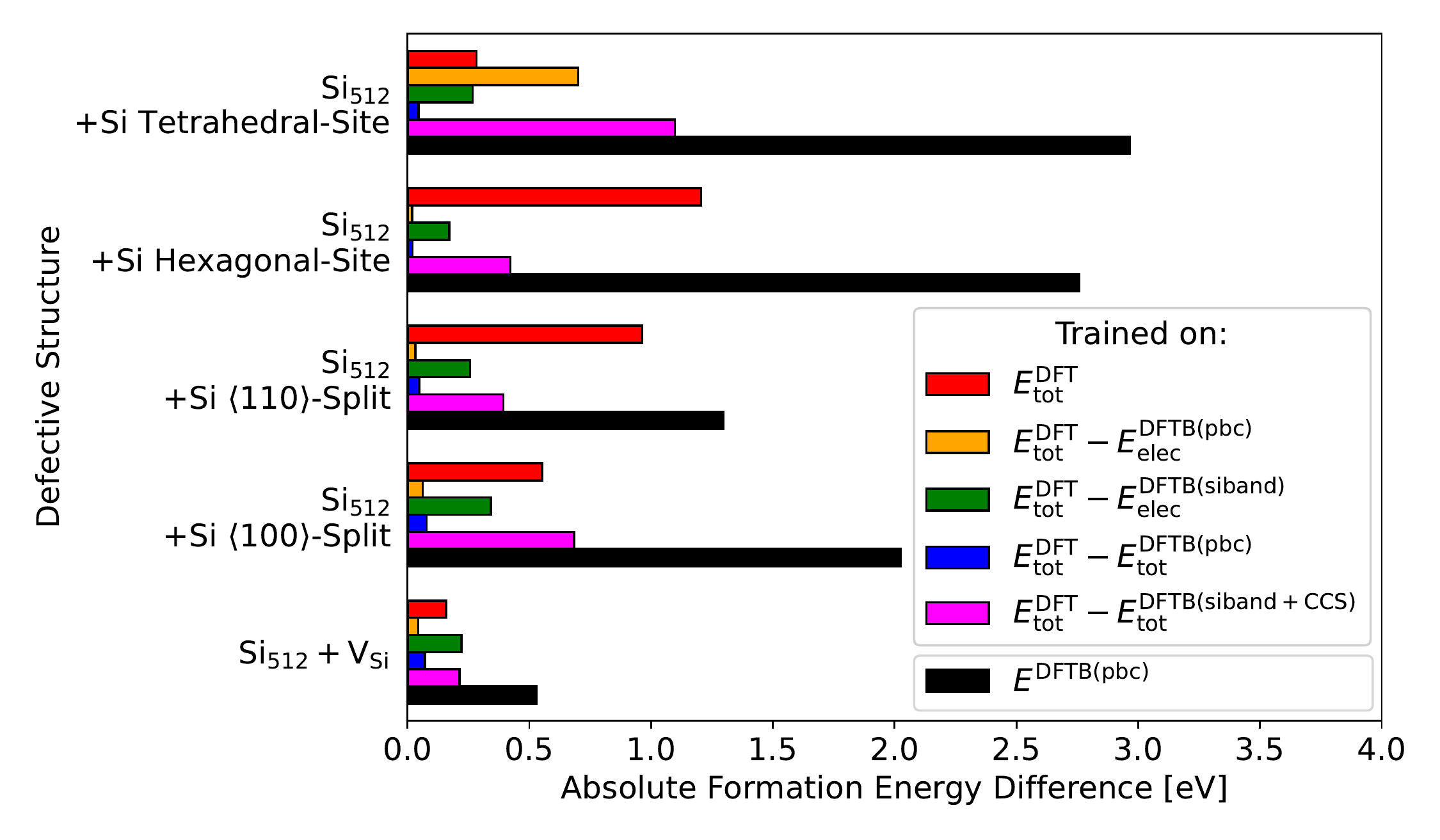}
\caption[DFTB Based DFT Reproduction of Formation Energies for Various Energy Terms as Training Targets]{Absolute formation energy difference for various silicon point defects and diamond bulk structure, w.r.t.\ GGA-PBE DFT. The bars correpsond to the objectives described in Section \ref{sec:objective}. Additionally, the results obtained with the original, uncorrected pbc set are also shown ($E^\mathrm{DFTB(pbc)}$). }
\label{fig:silicon-defect-dft-formen-scale}
\end{figure}

Finally, we further investigated the ANN performance, when calculating the formation energies of the five point defects in the large supercell. The formation energy $E_\mathrm{f}$ of a given defect was calculated as
\begin{align*}
  E_\mathrm{f} = E_\mathrm{tot} - \sum_i n_i \mu_i
\end{align*}
where $E_\mathrm{tot}$ is the total energy of the defective system, $n_i$ the number of atoms of a given atom type $i$ in the system and $\mu_i$ the chemical potential of an atom of type $i$ in a reference system. We have chosen the perfect diamond like bulk phase of silicon as reference system. Figure \ref{fig:silicon-defect-dft-formen-scale} illustrates the results, based on the ANN potentials. Since systematic errors may cancel out in the formation energy expression, the statements go beyond a plain accuracy comparison, in the sense that networks, whose error is more systematic should perform better. The formation energies also allow us to compare the performance of the ANN with the performance of the original, uncorrected pbc set for the same defect. It should be noted, that the correct description of the energetics of such point defects has not received special attention during the creation of the pbc-0-3 silicon parametrization~\cite{pbc}, consequently it performs significantly worse in predicting formation energies than its ANN corrected version, which shows the best performance. Interestingly, due to error cancellation, even the highly optimized siband set performs quite well in predicting the formation energies when corrected with the ANN.
\noindent

\section{Summary}

We have implemented a parallel Behler-Parrinello neural network in the open source \fnet{} software package. The software package offers a wide variety of features, including a truncated Xavier network parameter initialization (primarily intended to operate in combination with a sigmoidal activation function),
the capability to train on atomic or system-wide properties by using different loss functions and regularization techniques as well as an atomic force analysis. The parallel efficiency of the implementation was demonstrated by a realistic benchmark. Further key extensions are planned in the future, such as support for a mixture of atomic and system-wide target properties and library interfacing.

Using the \fnet{} software package, we have explored various possibilities to combine ANNs with the DFTB method in order to obtain DFT total energies and formation energies.
We have demonstrated, that by training an ANN to predict the difference between the total energies calculated by DFT and DFTB, one can achieve near DFT-accuracy, when calculating the formation energies of some selected point defects in silicon, at the computational cost of the much cheaper DFTB method. By training our ANN on small supercells only and subsequently evaluating its performance on significantly larger supercells, we could demonstrate that in contrast to the direct prediction of the DFT energies by an ANN, the combined DFTB+ANN approach features a significantly improved transferability. Although the computational cost of this combined method is higher as if only a pure ANN had to be evaluated, its superior transferability makes it nevertheless a promising approach for materials simulations. So far, our research was limited to single-point calculations, future studies will therefore plan to target molecular dynamic simulations.

\section*{Supplementary Material}
See supplementary material for exemplary Human-friendly Structured Data (HSD) input listings, as well as the basic usage of the \emph{Fortformat} Python layer for generating datasets and extracting results.

\begin{acknowledgments}
We would like to acknowledge the Swedish e-science initiative eSSENCE, the Swedish national infrastructure for computing (SNIC), and the Swedish research council (VR). The simulations were performed on the HPC cluster \emph{Aether} at the University of Bremen, financed by DFG within the scope of the Excellence Initiative.
\end{acknowledgments}

\section*{Data Availability Statement}
All datasets employed to obtain application related findings of this study are openly available at \url{https://doi.org/10.5281/zenodo.5969907}.

\appendix

\section{Analytical Expression for ASCF Derivatives}\label{sec:appendix-acsf-prime}
The derivatives of the radial ACSFs w.r.t\ atomic positions $\bm{R}$ are given by: 
\begin{subequations}
   \begin{align}
    \frac{\partial G_i^{1, Z_1}} {\partial R_{k \neq i}^{\alpha}} = & f_{\text{c}, ik}' \hat{R}_{ik}^{\alpha}
    \\
    \frac{\partial G_i^{2, Z_1}} {\partial R_{k \neq i}^{\alpha} } = & 
     \big(2\eta(R_s - R_{ij}) f_{\text{c}, ik} + f_{\text{c}, ik}' \big) e^{-\eta(R_{ij} - R_s)^2} \hat{R}_{ik}^{\alpha}
    \\
    \frac{\partial G_i^{3, Z_1}} {\partial R_{k \neq i}^{\alpha}} = & 
     \big(-\kappa \sin(\kappa R_{ik}) f_{\text{c},ik} + \cos(\kappa R_{ik}) f_{\text{c},ik}' \big) \hat{R}_{ik}^{\alpha}
   \end{align}
\end{subequations}
Here, $\hat{R}_{ab}^{\mu}$ is the $\mu$-th component of the unit vector from atom $a$ to atom $b$ and $R_{ab}$ is the distance between them. 
Moreover, $f_{\text{c}, ab} \equiv f_{\text{c}} (R_{ab})$, where $f_{\text{c}}$ is the cutoff function introduced by eq.\,\eqref{eq:f_cutoff} and $f_{\text{c}}'$ its first derivative:
\begin{equation}
    f_{\text{c}}'(r) = -\frac{\pi}{2R_c} \sin \left( \frac{\pi r}{R_c} \right)
\end{equation} 
The derivative of the $G^4$ function  w.r.t.\ $R_k^{\alpha}$ reads:
\begin{equation}\label{eq:G4_deriv}
  \begin{split}
     \frac{\partial G_i^{4, Z_1 Z_2}}{\partial R_{k \neq i}^{\alpha}} 
       = \, 2^{1-\xi} & \, \delta_{Z_k, Z_1} \\ 
       \times \bigg\{ \sum_{j \neq i}^{|Z_2|} f_{\text{c} \eta; ij}^{}& \, \mathcal{A}_{ijk}^{\xi-1} (\lambda) 
          \Big[ \lambda \, \xi \, f_{\text{c}, \eta; ik}^{} f_{\text{c} \eta; jk}^{}  \sum_{\beta} T_{ik}^{\alpha \beta} \hat{R}_{ij}^{\beta} \\
          + \mathcal{A}_{ijk}^{} (\lambda) & \big( f_{\text{c} \eta; ik}' f_{\text{c} \eta; jk}^{} \hat{R}_{ik}^{\alpha} + f_{\text{c} \eta; jk}' f_{\text{c} \eta; ik}^{} \hat{R}_{jk}^{\alpha} \big) \Big] 
        \bigg\} 
        \\[10pt]
        +\;  2^{1-\xi}\, \delta_{Z_k, Z_2} & \, \Big\{ Z_2 \rightarrow Z_1 \Big\},
  \end{split}
\end{equation}
Here, $\{ Z_2 \rightarrow Z_1 \}$ indicates that the preceding expression in curly brackets is to be repeated, with $Z_2$ substituted by $Z_1$, $f_{\text{c}\eta} (r) \equiv e^{-\eta r^2} f_{\text{c}} (r)$ is the exponentially damped cutoff function 
($f_{\text{c}\eta; ab} \equiv f_{\text{c}\eta} (R_{ab})$),
while $\mathcal{A}_{ijk} (\lambda)$ and $T_{ab}^{\mu \nu}$ are defined as:
\begin{subequations}
   \begin{align}
       \mathcal{A}_{ijk} (\lambda) \equiv & \; 1 + \lambda \hat{\bm{R}}_{ij} \cdot \hat{\bm{R}}_{ik}
       \\
       T_{ab}^{\mu \nu} \equiv & \; \frac{1}{R_{ab}} \big( \delta_{\mu\nu} - \hat{R}_{ab}^{\mu} \hat{R}_{ab}^{\nu} \big) 
   \end{align}
\end{subequations}
The formula for the derivative of the $G^5$ function can be obtained directly from \eqref{eq:G4_deriv} by treating $f_{\text{c} \eta;jk}$ as a constant equal to one.
Finally, all the preceding expressions are valid for taking the derivatives of the $G_i$ functions w.r.t\ positions of other atoms (i.e., whose index is not $i$).
Due to the translational invariance, the derivative of any $G_i$-function w.r.t.\ the position of atom $i$ can be obtained from its derivatives w.r.t.\ other atoms, following the expression below:
\begin{equation}
   \frac{\partial G_i^{}}{\partial R_i^{\alpha}} 
   =
   -\sum_{j \neq i} \frac {\partial G_i^{}}{\partial R_j^{\alpha}} 
\end{equation}

\bibliography{references}

\begin{thebibliography}{58}%
\makeatletter
\providecommand \@ifxundefined [1]{%
 \@ifx{#1\undefined}
}%
\providecommand \@ifnum [1]{%
 \ifnum #1\expandafter \@firstoftwo
 \else \expandafter \@secondoftwo
 \fi
}%
\providecommand \@ifx [1]{%
 \ifx #1\expandafter \@firstoftwo
 \else \expandafter \@secondoftwo
 \fi
}%
\providecommand \natexlab [1]{#1}%
\providecommand \enquote  [1]{``#1''}%
\providecommand \bibnamefont  [1]{#1}%
\providecommand \bibfnamefont [1]{#1}%
\providecommand \citenamefont [1]{#1}%
\providecommand \href@noop [0]{\@secondoftwo}%
\providecommand \href [0]{\begingroup \@sanitize@url \@href}%
\providecommand \@href[1]{\@@startlink{#1}\@@href}%
\providecommand \@@href[1]{\endgroup#1\@@endlink}%
\providecommand \@sanitize@url [0]{\catcode `\\12\catcode `\$12\catcode
  `\&12\catcode `\#12\catcode `\^12\catcode `\_12\catcode `\%12\relax}%
\providecommand \@@startlink[1]{}%
\providecommand \@@endlink[0]{}%
\providecommand \url  [0]{\begingroup\@sanitize@url \@url }%
\providecommand \@url [1]{\endgroup\@href {#1}{\urlprefix }}%
\providecommand \urlprefix  [0]{URL }%
\providecommand \Eprint [0]{\href }%
\providecommand \doibase [0]{http://dx.doi.org/}%
\providecommand \selectlanguage [0]{\@gobble}%
\providecommand \bibinfo  [0]{\@secondoftwo}%
\providecommand \bibfield  [0]{\@secondoftwo}%
\providecommand \translation [1]{[#1]}%
\providecommand \BibitemOpen [0]{}%
\providecommand \bibitemStop [0]{}%
\providecommand \bibitemNoStop [0]{.\EOS\space}%
\providecommand \EOS [0]{\spacefactor3000\relax}%
\providecommand \BibitemShut  [1]{\csname bibitem#1\endcsname}%
\let\auto@bib@innerbib\@empty
\bibitem [{\citenamefont {McCulloch}\ and\ \citenamefont
  {Pitts}(1943)}]{cullochpitts}%
  \BibitemOpen
  \bibfield  {author} {\bibinfo {author} {\bibfnamefont {W.~S.}\ \bibnamefont
  {McCulloch}}\ and\ \bibinfo {author} {\bibfnamefont {W.}~\bibnamefont
  {Pitts}},\ }\href {\doibase 10.1007/BF02478259} {\bibfield  {journal}
  {\bibinfo  {journal} {The bulletin of mathematical biophysics}\ }\textbf
  {\bibinfo {volume} {5}},\ \bibinfo {pages} {115} (\bibinfo {year}
  {1943})}\BibitemShut {NoStop}%
\bibitem [{\citenamefont {Hornik}, \citenamefont {Stinchcombe},\ and\
  \citenamefont {White}(1989)}]{uniapprox}%
  \BibitemOpen
  \bibfield  {author} {\bibinfo {author} {\bibfnamefont {K.}~\bibnamefont
  {Hornik}}, \bibinfo {author} {\bibfnamefont {M.}~\bibnamefont {Stinchcombe}},
  \ and\ \bibinfo {author} {\bibfnamefont {H.}~\bibnamefont {White}},\ }\href
  {\doibase 10.1016/0893-6080(89)90020-8} {\bibfield  {journal} {\bibinfo
  {journal} {Neural Networks}\ }\textbf {\bibinfo {volume} {2}},\ \bibinfo
  {pages} {359} (\bibinfo {year} {1989})}\BibitemShut {NoStop}%
\bibitem [{\citenamefont {Egmont-Petersen}, \citenamefont {{de Ridder}},\ and\
  \citenamefont {Handels}(2002)}]{ann_image_processing}%
  \BibitemOpen
  \bibfield  {author} {\bibinfo {author} {\bibfnamefont {M.}~\bibnamefont
  {Egmont-Petersen}}, \bibinfo {author} {\bibfnamefont {D.}~\bibnamefont {{de
  Ridder}}}, \ and\ \bibinfo {author} {\bibfnamefont {H.}~\bibnamefont
  {Handels}},\ }\href {\doibase 10.1016/S0031-3203(01)00178-9} {\bibfield
  {journal} {\bibinfo  {journal} {Lect Notes Comput Sc}\ }\textbf {\bibinfo
  {volume} {35}},\ \bibinfo {pages} {2279} (\bibinfo {year}
  {2002})}\BibitemShut {NoStop}%
\bibitem [{\citenamefont {Dahl}\ \emph {et~al.}(2012)\citenamefont {Dahl},
  \citenamefont {Yu}, \citenamefont {Deng},\ and\ \citenamefont
  {Acero}}]{ann_speech_recog1}%
  \BibitemOpen
  \bibfield  {author} {\bibinfo {author} {\bibfnamefont {G.~E.}\ \bibnamefont
  {Dahl}}, \bibinfo {author} {\bibfnamefont {D.}~\bibnamefont {Yu}}, \bibinfo
  {author} {\bibfnamefont {L.}~\bibnamefont {Deng}}, \ and\ \bibinfo {author}
  {\bibfnamefont {A.}~\bibnamefont {Acero}},\ }\href {\doibase
  10.1109/TASL.2011.2134090} {\bibfield  {journal} {\bibinfo  {journal} {IEEE
  Transactions on Audio, Speech, and Language Processing}\ }\textbf {\bibinfo
  {volume} {20}},\ \bibinfo {pages} {30} (\bibinfo {year} {2012})}\BibitemShut
  {NoStop}%
\bibitem [{\citenamefont {Sainath}\ \emph {et~al.}(2015)\citenamefont
  {Sainath}, \citenamefont {Vinyals}, \citenamefont {Senior},\ and\
  \citenamefont {Sak}}]{ann_speech_recog2}%
  \BibitemOpen
  \bibfield  {author} {\bibinfo {author} {\bibfnamefont {T.~N.}\ \bibnamefont
  {Sainath}}, \bibinfo {author} {\bibfnamefont {O.}~\bibnamefont {Vinyals}},
  \bibinfo {author} {\bibfnamefont {A.}~\bibnamefont {Senior}}, \ and\ \bibinfo
  {author} {\bibfnamefont {H.}~\bibnamefont {Sak}},\ }in\ \href {\doibase
  10.1109/ICASSP.2015.7178838} {\emph {\bibinfo {booktitle} {2015 IEEE
  International Conference on Acoustics, Speech and Signal Processing
  (ICASSP)}}}\ (\bibinfo {year} {2015})\ pp.\ \bibinfo {pages}
  {4580--4584}\BibitemShut {NoStop}%
\bibitem [{\citenamefont {Trentin}\ and\ \citenamefont
  {Gori}(2001)}]{ann_speech_recog3}%
  \BibitemOpen
  \bibfield  {author} {\bibinfo {author} {\bibfnamefont {E.}~\bibnamefont
  {Trentin}}\ and\ \bibinfo {author} {\bibfnamefont {M.}~\bibnamefont {Gori}},\
  }\href {\doibase 10.1016/S0925-2312(00)00308-8} {\bibfield  {journal}
  {\bibinfo  {journal} {Neurocomputing}\ }\textbf {\bibinfo {volume} {37}},\
  \bibinfo {pages} {91} (\bibinfo {year} {2001})}\BibitemShut {NoStop}%
\bibitem [{\citenamefont {Behler}\ and\ \citenamefont
  {Parrinello}(2007)}]{bpnn}%
  \BibitemOpen
  \bibfield  {author} {\bibinfo {author} {\bibfnamefont {J.}~\bibnamefont
  {Behler}}\ and\ \bibinfo {author} {\bibfnamefont {M.}~\bibnamefont
  {Parrinello}},\ }\href {\doibase 10.1103/PhysRevLett.98.146401} {\bibfield
  {journal} {\bibinfo  {journal} {Phys. Rev. Lett.}\ }\textbf {\bibinfo
  {volume} {98}},\ \bibinfo {pages} {146401} (\bibinfo {year}
  {2007})}\BibitemShut {NoStop}%
\bibitem [{\citenamefont {Goodfellow}, \citenamefont {Bengio},\ and\
  \citenamefont {Courville}(2016)}]{deeplbook}%
  \BibitemOpen
  \bibfield  {author} {\bibinfo {author} {\bibfnamefont {I.}~\bibnamefont
  {Goodfellow}}, \bibinfo {author} {\bibfnamefont {Y.}~\bibnamefont {Bengio}},
  \ and\ \bibinfo {author} {\bibfnamefont {A.}~\bibnamefont {Courville}},\
  }\href {http://www.deeplearningbook.org} {\emph {\bibinfo {title} {{Deep
  Learning}}}}\ (\bibinfo  {publisher} {MIT Press},\ \bibinfo {year}
  {2016})\BibitemShut {NoStop}%
\bibitem [{\citenamefont {Hohenberg}\ and\ \citenamefont {Kohn}(1964)}]{DFT}%
  \BibitemOpen
  \bibfield  {author} {\bibinfo {author} {\bibfnamefont {P.}~\bibnamefont
  {Hohenberg}}\ and\ \bibinfo {author} {\bibfnamefont {W.}~\bibnamefont
  {Kohn}},\ }\href {\doibase 10.1103/PhysRev.136.B864} {\bibfield  {journal}
  {\bibinfo  {journal} {Phys. Rev.}\ }\textbf {\bibinfo {volume} {136}},\
  \bibinfo {pages} {B864} (\bibinfo {year} {1964})}\BibitemShut {NoStop}%
\bibitem [{\citenamefont {Hourahine}\ \emph {et~al.}(2020)\citenamefont
  {Hourahine}, \citenamefont {Aradi}, \citenamefont {Blum}, \citenamefont
  {Bonaf{\'{e}}}, \citenamefont {Buccheri}, \citenamefont {Camacho},
  \citenamefont {Cevallos}, \citenamefont {Deshaye}, \citenamefont
  {Dumitric{\u{a}}}, \citenamefont {Dominguez}, \citenamefont {Ehlert},
  \citenamefont {Elstner}, \citenamefont {van~der Heide}, \citenamefont
  {Hermann}, \citenamefont {Irle}, \citenamefont {Kranz}, \citenamefont
  {Köhler}, \citenamefont {Kowalczyk}, \citenamefont {Kuba{\v{r}}},
  \citenamefont {Lee}, \citenamefont {Lutsker}, \citenamefont {Maurer},
  \citenamefont {Min}, \citenamefont {Mitchell}, \citenamefont {Negre},
  \citenamefont {Niehaus}, \citenamefont {Niklasson}, \citenamefont {Page},
  \citenamefont {Pecchia}, \citenamefont {Penazzi}, \citenamefont {Persson},
  \citenamefont {{\v{R}}ez{\'{a}}{\v{c}}}, \citenamefont {S{\'{a}}nchez},
  \citenamefont {Sternberg}, \citenamefont {Stöhr}, \citenamefont
  {Stuckenberg}, \citenamefont {Tkatchenko}, \citenamefont {z.~Yu},\ and\
  \citenamefont {Frauenheim}}]{newdftb}%
  \BibitemOpen
  \bibfield  {author} {\bibinfo {author} {\bibfnamefont {B.}~\bibnamefont
  {Hourahine}}, \bibinfo {author} {\bibfnamefont {B.}~\bibnamefont {Aradi}},
  \bibinfo {author} {\bibfnamefont {V.}~\bibnamefont {Blum}}, \bibinfo {author}
  {\bibfnamefont {F.}~\bibnamefont {Bonaf{\'{e}}}}, \bibinfo {author}
  {\bibfnamefont {A.}~\bibnamefont {Buccheri}}, \bibinfo {author}
  {\bibfnamefont {C.}~\bibnamefont {Camacho}}, \bibinfo {author} {\bibfnamefont
  {C.}~\bibnamefont {Cevallos}}, \bibinfo {author} {\bibfnamefont {M.~Y.}\
  \bibnamefont {Deshaye}}, \bibinfo {author} {\bibfnamefont {T.}~\bibnamefont
  {Dumitric{\u{a}}}}, \bibinfo {author} {\bibfnamefont {A.}~\bibnamefont
  {Dominguez}}, \bibinfo {author} {\bibfnamefont {S.}~\bibnamefont {Ehlert}},
  \bibinfo {author} {\bibfnamefont {M.}~\bibnamefont {Elstner}}, \bibinfo
  {author} {\bibfnamefont {T.}~\bibnamefont {van~der Heide}}, \bibinfo {author}
  {\bibfnamefont {J.}~\bibnamefont {Hermann}}, \bibinfo {author} {\bibfnamefont
  {S.}~\bibnamefont {Irle}}, \bibinfo {author} {\bibfnamefont {J.~J.}\
  \bibnamefont {Kranz}}, \bibinfo {author} {\bibfnamefont {C.}~\bibnamefont
  {Köhler}}, \bibinfo {author} {\bibfnamefont {T.}~\bibnamefont {Kowalczyk}},
  \bibinfo {author} {\bibfnamefont {T.}~\bibnamefont {Kuba{\v{r}}}}, \bibinfo
  {author} {\bibfnamefont {I.~S.}\ \bibnamefont {Lee}}, \bibinfo {author}
  {\bibfnamefont {V.}~\bibnamefont {Lutsker}}, \bibinfo {author} {\bibfnamefont
  {R.~J.}\ \bibnamefont {Maurer}}, \bibinfo {author} {\bibfnamefont {S.~K.}\
  \bibnamefont {Min}}, \bibinfo {author} {\bibfnamefont {I.}~\bibnamefont
  {Mitchell}}, \bibinfo {author} {\bibfnamefont {C.}~\bibnamefont {Negre}},
  \bibinfo {author} {\bibfnamefont {T.~A.}\ \bibnamefont {Niehaus}}, \bibinfo
  {author} {\bibfnamefont {A.~M.~N.}\ \bibnamefont {Niklasson}}, \bibinfo
  {author} {\bibfnamefont {A.~J.}\ \bibnamefont {Page}}, \bibinfo {author}
  {\bibfnamefont {A.}~\bibnamefont {Pecchia}}, \bibinfo {author} {\bibfnamefont
  {G.}~\bibnamefont {Penazzi}}, \bibinfo {author} {\bibfnamefont {M.~P.}\
  \bibnamefont {Persson}}, \bibinfo {author} {\bibfnamefont {J.}~\bibnamefont
  {{\v{R}}ez{\'{a}}{\v{c}}}}, \bibinfo {author} {\bibfnamefont {C.~G.}\
  \bibnamefont {S{\'{a}}nchez}}, \bibinfo {author} {\bibfnamefont
  {M.}~\bibnamefont {Sternberg}}, \bibinfo {author} {\bibfnamefont
  {M.}~\bibnamefont {Stöhr}}, \bibinfo {author} {\bibfnamefont
  {F.}~\bibnamefont {Stuckenberg}}, \bibinfo {author} {\bibfnamefont
  {A.}~\bibnamefont {Tkatchenko}}, \bibinfo {author} {\bibfnamefont {V.~W.}\
  \bibnamefont {z.~Yu}}, \ and\ \bibinfo {author} {\bibfnamefont
  {T.}~\bibnamefont {Frauenheim}},\ }\href {\doibase 10.1063/1.5143190}
  {\bibfield  {journal} {\bibinfo  {journal} {J. Chem. Phys.}\ }\textbf
  {\bibinfo {volume} {152}},\ \bibinfo {pages} {124101} (\bibinfo {year}
  {2020})}\BibitemShut {NoStop}%
\bibitem [{\citenamefont {Blank}\ \emph {et~al.}(1995)\citenamefont {Blank},
  \citenamefont {Brown}, \citenamefont {Calhoun},\ and\ \citenamefont
  {Doren}}]{ffnn_first_gen}%
  \BibitemOpen
  \bibfield  {author} {\bibinfo {author} {\bibfnamefont {T.~B.}\ \bibnamefont
  {Blank}}, \bibinfo {author} {\bibfnamefont {S.~D.}\ \bibnamefont {Brown}},
  \bibinfo {author} {\bibfnamefont {A.~W.}\ \bibnamefont {Calhoun}}, \ and\
  \bibinfo {author} {\bibfnamefont {D.~J.}\ \bibnamefont {Doren}},\ }\href
  {\doibase 10.1063/1.469597} {\bibfield  {journal} {\bibinfo  {journal} {J.
  Chem. Phys.}\ }\textbf {\bibinfo {volume} {103}},\ \bibinfo {pages} {4129}
  (\bibinfo {year} {1995})}\BibitemShut {NoStop}%
\bibitem [{\citenamefont {Artrith}, \citenamefont {Morawietz},\ and\
  \citenamefont {Behler}(2011)}]{ffnn_third_gen1}%
  \BibitemOpen
  \bibfield  {author} {\bibinfo {author} {\bibfnamefont {N.}~\bibnamefont
  {Artrith}}, \bibinfo {author} {\bibfnamefont {T.}~\bibnamefont {Morawietz}},
  \ and\ \bibinfo {author} {\bibfnamefont {J.}~\bibnamefont {Behler}},\ }\href
  {\doibase 10.1103/PhysRevB.83.153101} {\bibfield  {journal} {\bibinfo
  {journal} {Phys. Rev. B}\ }\textbf {\bibinfo {volume} {83}},\ \bibinfo
  {pages} {153101} (\bibinfo {year} {2011})}\BibitemShut {NoStop}%
\bibitem [{\citenamefont {Artrith}, \citenamefont {Morawietz},\ and\
  \citenamefont {Behler}(2012)}]{ffnn_third_gen2}%
  \BibitemOpen
  \bibfield  {author} {\bibinfo {author} {\bibfnamefont {N.}~\bibnamefont
  {Artrith}}, \bibinfo {author} {\bibfnamefont {T.}~\bibnamefont {Morawietz}},
  \ and\ \bibinfo {author} {\bibfnamefont {J.}~\bibnamefont {Behler}},\ }\href
  {\doibase 10.1103/PhysRevB.86.079914} {\bibfield  {journal} {\bibinfo
  {journal} {Phys. Rev. B}\ }\textbf {\bibinfo {volume} {86}},\ \bibinfo
  {pages} {079914} (\bibinfo {year} {2012})}\BibitemShut {NoStop}%
\bibitem [{\citenamefont {Ko}\ \emph {et~al.}(2021)\citenamefont {Ko},
  \citenamefont {Finkler}, \citenamefont {Goedecker},\ and\ \citenamefont
  {Behler}}]{ffnn_fourth_gen}%
  \BibitemOpen
  \bibfield  {author} {\bibinfo {author} {\bibfnamefont {T.~W.}\ \bibnamefont
  {Ko}}, \bibinfo {author} {\bibfnamefont {J.~A.}\ \bibnamefont {Finkler}},
  \bibinfo {author} {\bibfnamefont {S.}~\bibnamefont {Goedecker}}, \ and\
  \bibinfo {author} {\bibfnamefont {J.}~\bibnamefont {Behler}},\ }\href
  {\doibase 10.1038/s41467-020-20427-2} {\bibfield  {journal} {\bibinfo
  {journal} {Nat Commun}\ }\textbf {\bibinfo {volume} {12}} (\bibinfo {year}
  {2021}),\ 10.1038/s41467-020-20427-2}\BibitemShut {NoStop}%
\bibitem [{\citenamefont {van~der Heide}(2021)}]{fortnet_zenodo}%
  \BibitemOpen
  \bibfield  {author} {\bibinfo {author} {\bibfnamefont {T.}~\bibnamefont
  {van~der Heide}},\ }\href {\doibase 10.5281/zenodo.5336841} {\enquote
  {\bibinfo {title} {Zenodo},}\ } (\bibinfo {year} {2021})\BibitemShut
  {NoStop}%
\bibitem [{\citenamefont {{The HDF Group}}(2010)}]{hdf5}%
  \BibitemOpen
  \bibfield  {author} {\bibinfo {author} {\bibnamefont {{The HDF Group}}},\
  }\href {http://www.hdfgroup.org/HDF5} {\enquote {\bibinfo {title}
  {{Hierarchical data format version 5}},}\ } (\bibinfo {year}
  {2000-2010})\BibitemShut {NoStop}%
\bibitem [{\citenamefont {Abadi}\ \emph {et~al.}(2015)\citenamefont {Abadi},
  \citenamefont {Agarwal}, \citenamefont {Barham}, \citenamefont {Brevdo},
  \citenamefont {Chen}, \citenamefont {Citro}, \citenamefont {Corrado},
  \citenamefont {Davis}, \citenamefont {Dean}, \citenamefont {Devin},
  \citenamefont {Ghemawat}, \citenamefont {Goodfellow}, \citenamefont {Harp},
  \citenamefont {Irving}, \citenamefont {Isard}, \citenamefont {Jia},
  \citenamefont {Jozefowicz}, \citenamefont {Kaiser}, \citenamefont {Kudlur},
  \citenamefont {Levenberg}, \citenamefont {Man\'{e}}, \citenamefont {Monga},
  \citenamefont {Moore}, \citenamefont {Murray}, \citenamefont {Olah},
  \citenamefont {Schuster}, \citenamefont {Shlens}, \citenamefont {Steiner},
  \citenamefont {Sutskever}, \citenamefont {Talwar}, \citenamefont {Tucker},
  \citenamefont {Vanhoucke}, \citenamefont {Vasudevan}, \citenamefont
  {Vi\'{e}gas}, \citenamefont {Vinyals}, \citenamefont {Warden}, \citenamefont
  {Wattenberg}, \citenamefont {Wicke}, \citenamefont {Yu},\ and\ \citenamefont
  {Zheng}}]{tensorflow}%
  \BibitemOpen
  \bibfield  {author} {\bibinfo {author} {\bibfnamefont {M.}~\bibnamefont
  {Abadi}}, \bibinfo {author} {\bibfnamefont {A.}~\bibnamefont {Agarwal}},
  \bibinfo {author} {\bibfnamefont {P.}~\bibnamefont {Barham}}, \bibinfo
  {author} {\bibfnamefont {E.}~\bibnamefont {Brevdo}}, \bibinfo {author}
  {\bibfnamefont {Z.}~\bibnamefont {Chen}}, \bibinfo {author} {\bibfnamefont
  {C.}~\bibnamefont {Citro}}, \bibinfo {author} {\bibfnamefont {G.~S.}\
  \bibnamefont {Corrado}}, \bibinfo {author} {\bibfnamefont {A.}~\bibnamefont
  {Davis}}, \bibinfo {author} {\bibfnamefont {J.}~\bibnamefont {Dean}},
  \bibinfo {author} {\bibfnamefont {M.}~\bibnamefont {Devin}}, \bibinfo
  {author} {\bibfnamefont {S.}~\bibnamefont {Ghemawat}}, \bibinfo {author}
  {\bibfnamefont {I.}~\bibnamefont {Goodfellow}}, \bibinfo {author}
  {\bibfnamefont {A.}~\bibnamefont {Harp}}, \bibinfo {author} {\bibfnamefont
  {G.}~\bibnamefont {Irving}}, \bibinfo {author} {\bibfnamefont
  {M.}~\bibnamefont {Isard}}, \bibinfo {author} {\bibfnamefont
  {Y.}~\bibnamefont {Jia}}, \bibinfo {author} {\bibfnamefont {R.}~\bibnamefont
  {Jozefowicz}}, \bibinfo {author} {\bibfnamefont {L.}~\bibnamefont {Kaiser}},
  \bibinfo {author} {\bibfnamefont {M.}~\bibnamefont {Kudlur}}, \bibinfo
  {author} {\bibfnamefont {J.}~\bibnamefont {Levenberg}}, \bibinfo {author}
  {\bibfnamefont {D.}~\bibnamefont {Man\'{e}}}, \bibinfo {author}
  {\bibfnamefont {R.}~\bibnamefont {Monga}}, \bibinfo {author} {\bibfnamefont
  {S.}~\bibnamefont {Moore}}, \bibinfo {author} {\bibfnamefont
  {D.}~\bibnamefont {Murray}}, \bibinfo {author} {\bibfnamefont
  {C.}~\bibnamefont {Olah}}, \bibinfo {author} {\bibfnamefont {M.}~\bibnamefont
  {Schuster}}, \bibinfo {author} {\bibfnamefont {J.}~\bibnamefont {Shlens}},
  \bibinfo {author} {\bibfnamefont {B.}~\bibnamefont {Steiner}}, \bibinfo
  {author} {\bibfnamefont {I.}~\bibnamefont {Sutskever}}, \bibinfo {author}
  {\bibfnamefont {K.}~\bibnamefont {Talwar}}, \bibinfo {author} {\bibfnamefont
  {P.}~\bibnamefont {Tucker}}, \bibinfo {author} {\bibfnamefont
  {V.}~\bibnamefont {Vanhoucke}}, \bibinfo {author} {\bibfnamefont
  {V.}~\bibnamefont {Vasudevan}}, \bibinfo {author} {\bibfnamefont
  {F.}~\bibnamefont {Vi\'{e}gas}}, \bibinfo {author} {\bibfnamefont
  {O.}~\bibnamefont {Vinyals}}, \bibinfo {author} {\bibfnamefont
  {P.}~\bibnamefont {Warden}}, \bibinfo {author} {\bibfnamefont
  {M.}~\bibnamefont {Wattenberg}}, \bibinfo {author} {\bibfnamefont
  {M.}~\bibnamefont {Wicke}}, \bibinfo {author} {\bibfnamefont
  {Y.}~\bibnamefont {Yu}}, \ and\ \bibinfo {author} {\bibfnamefont
  {X.}~\bibnamefont {Zheng}},\ }\href {https://www.tensorflow.org/} {\enquote
  {\bibinfo {title} {{{TensorFlow}: Large-Scale Machine Learning on
  Heterogeneous Systems}},}\ } (\bibinfo {year} {2015}),\ \bibinfo {note}
  {software available from tensorflow.org}\BibitemShut {NoStop}%
\bibitem [{\citenamefont {Chollet}\ \emph {et~al.}(2015)\citenamefont {Chollet}
  \emph {et~al.}}]{keras}%
  \BibitemOpen
  \bibfield  {author} {\bibinfo {author} {\bibfnamefont {F.}~\bibnamefont
  {Chollet}} \emph {et~al.},\ }\href@noop {} {\enquote {\bibinfo {title}
  {Keras},}\ }\bibinfo {howpublished} {\url{https://keras.io}} (\bibinfo {year}
  {2015})\BibitemShut {NoStop}%
\bibitem [{\citenamefont {Paszke}\ \emph {et~al.}(2017)\citenamefont {Paszke},
  \citenamefont {Gross}, \citenamefont {Chintala}, \citenamefont {Chanan},
  \citenamefont {Yang}, \citenamefont {DeVito}, \citenamefont {Lin},
  \citenamefont {Desmaison}, \citenamefont {Antiga},\ and\ \citenamefont
  {Lerer}}]{pytorch}%
  \BibitemOpen
  \bibfield  {author} {\bibinfo {author} {\bibfnamefont {A.}~\bibnamefont
  {Paszke}}, \bibinfo {author} {\bibfnamefont {S.}~\bibnamefont {Gross}},
  \bibinfo {author} {\bibfnamefont {S.}~\bibnamefont {Chintala}}, \bibinfo
  {author} {\bibfnamefont {G.}~\bibnamefont {Chanan}}, \bibinfo {author}
  {\bibfnamefont {E.}~\bibnamefont {Yang}}, \bibinfo {author} {\bibfnamefont
  {Z.}~\bibnamefont {DeVito}}, \bibinfo {author} {\bibfnamefont
  {Z.}~\bibnamefont {Lin}}, \bibinfo {author} {\bibfnamefont {A.}~\bibnamefont
  {Desmaison}}, \bibinfo {author} {\bibfnamefont {L.}~\bibnamefont {Antiga}}, \
  and\ \bibinfo {author} {\bibfnamefont {A.}~\bibnamefont {Lerer}},\ }in\
  \href@noop {} {\emph {\bibinfo {booktitle} {NIPS Autodiff Workshop}}}\
  (\bibinfo {year} {2017})\BibitemShut {NoStop}%
\bibitem [{\citenamefont {Pedregosa}\ \emph {et~al.}(2011)\citenamefont
  {Pedregosa}, \citenamefont {Varoquaux}, \citenamefont {Gramfort},
  \citenamefont {Michel}, \citenamefont {Thirion}, \citenamefont {Grisel},
  \citenamefont {Blondel}, \citenamefont {Prettenhofer}, \citenamefont {Weiss},
  \citenamefont {Dubourg}, \citenamefont {Vanderplas}, \citenamefont {Passos},
  \citenamefont {Cournapeau}, \citenamefont {Brucher}, \citenamefont {Perrot},\
  and\ \citenamefont {Duchesnay}}]{scikit-learn}%
  \BibitemOpen
  \bibfield  {author} {\bibinfo {author} {\bibfnamefont {F.}~\bibnamefont
  {Pedregosa}}, \bibinfo {author} {\bibfnamefont {G.}~\bibnamefont
  {Varoquaux}}, \bibinfo {author} {\bibfnamefont {A.}~\bibnamefont {Gramfort}},
  \bibinfo {author} {\bibfnamefont {V.}~\bibnamefont {Michel}}, \bibinfo
  {author} {\bibfnamefont {B.}~\bibnamefont {Thirion}}, \bibinfo {author}
  {\bibfnamefont {O.}~\bibnamefont {Grisel}}, \bibinfo {author} {\bibfnamefont
  {M.}~\bibnamefont {Blondel}}, \bibinfo {author} {\bibfnamefont
  {P.}~\bibnamefont {Prettenhofer}}, \bibinfo {author} {\bibfnamefont
  {R.}~\bibnamefont {Weiss}}, \bibinfo {author} {\bibfnamefont
  {V.}~\bibnamefont {Dubourg}}, \bibinfo {author} {\bibfnamefont
  {J.}~\bibnamefont {Vanderplas}}, \bibinfo {author} {\bibfnamefont
  {A.}~\bibnamefont {Passos}}, \bibinfo {author} {\bibfnamefont
  {D.}~\bibnamefont {Cournapeau}}, \bibinfo {author} {\bibfnamefont
  {M.}~\bibnamefont {Brucher}}, \bibinfo {author} {\bibfnamefont
  {M.}~\bibnamefont {Perrot}}, \ and\ \bibinfo {author} {\bibfnamefont
  {E.}~\bibnamefont {Duchesnay}},\ }\href@noop {} {\bibfield  {journal}
  {\bibinfo  {journal} {J Mach Learn Res}\ }\textbf {\bibinfo {volume} {12}},\
  \bibinfo {pages} {2825} (\bibinfo {year} {2011})}\BibitemShut {NoStop}%
\bibitem [{\citenamefont {Curcic}(2019)}]{neural-fortran}%
  \BibitemOpen
  \bibfield  {author} {\bibinfo {author} {\bibfnamefont {M.}~\bibnamefont
  {Curcic}},\ }\href {http://arxiv.org/abs/1902.06714} {\bibfield  {journal}
  {\bibinfo  {journal} {CoRR}\ }\textbf {\bibinfo {volume} {abs/1902.06714}}
  (\bibinfo {year} {2019})},\ \Eprint {http://arxiv.org/abs/1902.06714}
  {1902.06714} \BibitemShut {NoStop}%
\bibitem [{\citenamefont {Ammothum~Kandy}\ \emph {et~al.}(2021)\citenamefont
  {Ammothum~Kandy}, \citenamefont {Wadbro}, \citenamefont {Aradi},
  \citenamefont {Broqvist},\ and\ \citenamefont {Kullgren}}]{ccs}%
  \BibitemOpen
  \bibfield  {author} {\bibinfo {author} {\bibfnamefont {A.~K.}\ \bibnamefont
  {Ammothum~Kandy}}, \bibinfo {author} {\bibfnamefont {E.}~\bibnamefont
  {Wadbro}}, \bibinfo {author} {\bibfnamefont {B.}~\bibnamefont {Aradi}},
  \bibinfo {author} {\bibfnamefont {P.}~\bibnamefont {Broqvist}}, \ and\
  \bibinfo {author} {\bibfnamefont {J.}~\bibnamefont {Kullgren}},\ }\href
  {\doibase 10.1021/acs.jctc.0c01156} {\bibfield  {journal} {\bibinfo
  {journal} {J. Chem. Theory Comput.}\ }\textbf {\bibinfo {volume} {17}},\
  \bibinfo {pages} {1771} (\bibinfo {year} {2021})},\ \bibinfo {note} {pMID:
  33606527}\BibitemShut {NoStop}%
\bibitem [{\citenamefont {Perdew}, \citenamefont {Burke},\ and\ \citenamefont
  {Ernzerhof}(1996)}]{gga-pbe}%
  \BibitemOpen
  \bibfield  {author} {\bibinfo {author} {\bibfnamefont {J.~P.}\ \bibnamefont
  {Perdew}}, \bibinfo {author} {\bibfnamefont {K.}~\bibnamefont {Burke}}, \
  and\ \bibinfo {author} {\bibfnamefont {M.}~\bibnamefont {Ernzerhof}},\ }\href
  {\doibase 10.1103/PhysRevLett.77.3865} {\bibfield  {journal} {\bibinfo
  {journal} {Phys. Rev. Lett.}\ }\textbf {\bibinfo {volume} {77}},\ \bibinfo
  {pages} {3865} (\bibinfo {year} {1996})}\BibitemShut {NoStop}%
\bibitem [{\citenamefont {Kruse}\ \emph {et~al.}(2016)\citenamefont {Kruse},
  \citenamefont {Borgelt}, \citenamefont {Braune}, \citenamefont {Mostaghim},\
  and\ \citenamefont {Steinbrecher}}]{cieng}%
  \BibitemOpen
  \bibfield  {author} {\bibinfo {author} {\bibfnamefont {R.}~\bibnamefont
  {Kruse}}, \bibinfo {author} {\bibfnamefont {C.}~\bibnamefont {Borgelt}},
  \bibinfo {author} {\bibfnamefont {C.}~\bibnamefont {Braune}}, \bibinfo
  {author} {\bibfnamefont {S.}~\bibnamefont {Mostaghim}}, \ and\ \bibinfo
  {author} {\bibfnamefont {M.}~\bibnamefont {Steinbrecher}},\ }\href {\doibase
  10.1007/978-1-4471-7296-3} {\emph {\bibinfo {title} {{Computational
  Intelligence: A Methodological Introduction}}}},\ \bibinfo {edition} {two}\
  ed.\ (\bibinfo  {publisher} {Springer-Verlag London},\ \bibinfo {year}
  {2016})\BibitemShut {NoStop}%
\bibitem [{\citenamefont {Rosenblatt}(1958)}]{perceptron}%
  \BibitemOpen
  \bibfield  {author} {\bibinfo {author} {\bibfnamefont {F.}~\bibnamefont
  {Rosenblatt}},\ }\href@noop {} {\bibfield  {journal} {\bibinfo  {journal}
  {Psychol Rev}\ ,\ \bibinfo {pages} {65}} (\bibinfo {year}
  {1958})}\BibitemShut {NoStop}%
\bibitem [{\citenamefont {Nielsen}(2015)}]{bpropderivation}%
  \BibitemOpen
  \bibfield  {author} {\bibinfo {author} {\bibfnamefont {M.~A.}\ \bibnamefont
  {Nielsen}},\ }\href@noop {} {\emph {\bibinfo {title} {{Neural Networks and
  Deep Learning}}}}\ (\bibinfo  {publisher} {Determination Press},\ \bibinfo
  {year} {2015})\BibitemShut {NoStop}%
\bibitem [{\citenamefont {Behler}(2011)}]{acsf}%
  \BibitemOpen
  \bibfield  {author} {\bibinfo {author} {\bibfnamefont {J.}~\bibnamefont
  {Behler}},\ }\href {\doibase 10.1063/1.3553717} {\bibfield  {journal}
  {\bibinfo  {journal} {J. Chem. Phys.}\ }\textbf {\bibinfo {volume} {134}},\
  \bibinfo {pages} {074106} (\bibinfo {year} {2011})}\BibitemShut {NoStop}%
\bibitem [{Note1()}]{Note1}%
  \BibitemOpen
  \bibinfo {note} {See \protect \url
  {https://fortnet.readthedocs.io}}\BibitemShut {NoStop}%
\bibitem [{\citenamefont {Gastegger}\ \emph {et~al.}(2018)\citenamefont
  {Gastegger}, \citenamefont {Schwiedrzik}, \citenamefont {Bittermann},
  \citenamefont {Berzsenyi},\ and\ \citenamefont {Marquetand}}]{wACSF}%
  \BibitemOpen
  \bibfield  {author} {\bibinfo {author} {\bibfnamefont {M.}~\bibnamefont
  {Gastegger}}, \bibinfo {author} {\bibfnamefont {L.}~\bibnamefont
  {Schwiedrzik}}, \bibinfo {author} {\bibfnamefont {M.}~\bibnamefont
  {Bittermann}}, \bibinfo {author} {\bibfnamefont {F.}~\bibnamefont
  {Berzsenyi}}, \ and\ \bibinfo {author} {\bibfnamefont {P.}~\bibnamefont
  {Marquetand}},\ }\href {\doibase 10.1063/1.5019667} {\bibfield  {journal}
  {\bibinfo  {journal} {The Journal of Chemical Physics}\ }\textbf {\bibinfo
  {volume} {148}},\ \bibinfo {pages} {241709} (\bibinfo {year}
  {2018})}\BibitemShut {NoStop}%
\bibitem [{\citenamefont {Glorot}\ and\ \citenamefont
  {Bengio}(2010)}]{xavierinit}%
  \BibitemOpen
  \bibfield  {author} {\bibinfo {author} {\bibfnamefont {X.}~\bibnamefont
  {Glorot}}\ and\ \bibinfo {author} {\bibfnamefont {Y.}~\bibnamefont
  {Bengio}},\ }in\ \href {http://proceedings.mlr.press/v9/glorot10a.html}
  {\emph {\bibinfo {booktitle} {Proceedings of the Thirteenth International
  Conference on Artificial Intelligence and Statistics}}},\ \bibinfo {series}
  {Proceedings of Machine Learning Research}, Vol.~\bibinfo {volume} {9},\
  \bibinfo {editor} {edited by\ \bibinfo {editor} {\bibfnamefont {Y.~W.}\
  \bibnamefont {Teh}}\ and\ \bibinfo {editor} {\bibfnamefont {M.}~\bibnamefont
  {Titterington}}}\ (\bibinfo  {publisher} {PMLR},\ \bibinfo {address} {Chia
  Laguna Resort, Sardinia, Italy},\ \bibinfo {year} {2010})\ pp.\ \bibinfo
  {pages} {249--256}\BibitemShut {NoStop}%
\bibitem [{\citenamefont {Marsaglia}\ and\ \citenamefont
  {Zaman}(1991)}]{ranlux1}%
  \BibitemOpen
  \bibfield  {author} {\bibinfo {author} {\bibfnamefont {G.}~\bibnamefont
  {Marsaglia}}\ and\ \bibinfo {author} {\bibfnamefont {A.}~\bibnamefont
  {Zaman}},\ }\href {\doibase 10.1214/aoap/1177005878} {\bibfield  {journal}
  {\bibinfo  {journal} {Ann. Appl. Probab.}\ }\textbf {\bibinfo {volume} {1}},\
  \bibinfo {pages} {462} (\bibinfo {year} {1991})}\BibitemShut {NoStop}%
\bibitem [{\citenamefont {Lüscher}(1994)}]{ranlux2}%
  \BibitemOpen
  \bibfield  {author} {\bibinfo {author} {\bibfnamefont {M.}~\bibnamefont
  {Lüscher}},\ }\href {\doibase 10.1016/0010-4655(94)90232-1} {\bibfield
  {journal} {\bibinfo  {journal} {Comput. Phys. Commun.}\ }\textbf {\bibinfo
  {volume} {79}},\ \bibinfo {pages} {100} (\bibinfo {year} {1994})}\BibitemShut
  {NoStop}%
\bibitem [{\citenamefont {James}(1994)}]{ranlux3}%
  \BibitemOpen
  \bibfield  {author} {\bibinfo {author} {\bibfnamefont {F.}~\bibnamefont
  {James}},\ }\href {\doibase 10.1016/0010-4655(94)90233-X} {\bibfield
  {journal} {\bibinfo  {journal} {Comput. Phys. Commun.}\ }\textbf {\bibinfo
  {volume} {79}},\ \bibinfo {pages} {111} (\bibinfo {year} {1994})}\BibitemShut
  {NoStop}%
\bibitem [{\citenamefont {Werbos}(1974)}]{bprophist}%
  \BibitemOpen
  \bibfield  {author} {\bibinfo {author} {\bibfnamefont {P.}~\bibnamefont
  {Werbos}},\ }\href@noop {} {\bibfield  {journal} {\bibinfo  {journal} {Ph. D.
  dissertation, Harvard University}\ } (\bibinfo {year} {1974})}\BibitemShut
  {NoStop}%
\bibitem [{\citenamefont {Rumelhart}, \citenamefont {Hinton},\ and\
  \citenamefont {Williams}(1986)}]{bprop}%
  \BibitemOpen
  \bibfield  {author} {\bibinfo {author} {\bibfnamefont {D.~E.}\ \bibnamefont
  {Rumelhart}}, \bibinfo {author} {\bibfnamefont {G.~E.}\ \bibnamefont
  {Hinton}}, \ and\ \bibinfo {author} {\bibfnamefont {R.~J.}\ \bibnamefont
  {Williams}},\ }\href {\doibase 10.1038/323533a0} {\bibfield  {journal}
  {\bibinfo  {journal} {Nature}\ }\textbf {\bibinfo {volume} {323}},\ \bibinfo
  {pages} {533} (\bibinfo {year} {1986})}\BibitemShut {NoStop}%
\bibitem [{\citenamefont {Nair}\ and\ \citenamefont {Hinton}(2010)}]{relu}%
  \BibitemOpen
  \bibfield  {author} {\bibinfo {author} {\bibfnamefont {V.}~\bibnamefont
  {Nair}}\ and\ \bibinfo {author} {\bibfnamefont {G.~E.}\ \bibnamefont
  {Hinton}},\ }in\ \href {https://icml.cc/Conferences/2010/papers/432.pdf}
  {\emph {\bibinfo {booktitle} {ICML}}}\ (\bibinfo {year} {2010})\ pp.\
  \bibinfo {pages} {807--814}\BibitemShut {NoStop}%
\bibitem [{\citenamefont {Dugas}\ \emph {et~al.}(2001)\citenamefont {Dugas},
  \citenamefont {Bengio}, \citenamefont {B{\'e}lisle}, \citenamefont {Nadeau},\
  and\ \citenamefont {Garcia}}]{softplus}%
  \BibitemOpen
  \bibfield  {author} {\bibinfo {author} {\bibfnamefont {C.}~\bibnamefont
  {Dugas}}, \bibinfo {author} {\bibfnamefont {Y.}~\bibnamefont {Bengio}},
  \bibinfo {author} {\bibfnamefont {F.}~\bibnamefont {B{\'e}lisle}}, \bibinfo
  {author} {\bibfnamefont {C.}~\bibnamefont {Nadeau}}, \ and\ \bibinfo {author}
  {\bibfnamefont {R.}~\bibnamefont {Garcia}},\ }\href@noop {} {\bibfield
  {journal} {\bibinfo  {journal} {Advances in neural information processing
  systems}\ ,\ \bibinfo {pages} {472}} (\bibinfo {year} {2001})}\BibitemShut
  {NoStop}%
\bibitem [{\citenamefont {Krizhevsky}, \citenamefont {Sutskever},\ and\
  \citenamefont {Hinton}(2012)}]{relusuccess}%
  \BibitemOpen
  \bibfield  {author} {\bibinfo {author} {\bibfnamefont {A.}~\bibnamefont
  {Krizhevsky}}, \bibinfo {author} {\bibfnamefont {I.}~\bibnamefont
  {Sutskever}}, \ and\ \bibinfo {author} {\bibfnamefont {G.~E.}\ \bibnamefont
  {Hinton}},\ }\href@noop {} {\bibfield  {journal} {\bibinfo  {journal}
  {Advances in neural information processing systems}\ }\textbf {\bibinfo
  {volume} {25}},\ \bibinfo {pages} {1097} (\bibinfo {year}
  {2012})}\BibitemShut {NoStop}%
\bibitem [{\citenamefont {Tibshirani}(1996)}]{lasso}%
  \BibitemOpen
  \bibfield  {author} {\bibinfo {author} {\bibfnamefont {R.}~\bibnamefont
  {Tibshirani}},\ }\href {\doibase 10.1111/j.2517-6161.1996.tb02080.x}
  {\bibfield  {journal} {\bibinfo  {journal} {Journal of the Royal Statistical
  Society: Series B (Methodological)}\ }\textbf {\bibinfo {volume} {58}},\
  \bibinfo {pages} {267} (\bibinfo {year} {1996})}\BibitemShut {NoStop}%
\bibitem [{\citenamefont {Hoerl}\ and\ \citenamefont {Kennard}(1988)}]{ridge}%
  \BibitemOpen
  \bibfield  {author} {\bibinfo {author} {\bibfnamefont {A.}~\bibnamefont
  {Hoerl}}\ and\ \bibinfo {author} {\bibfnamefont {R.}~\bibnamefont
  {Kennard}},\ }\href@noop {} {\enquote {\bibinfo {title} {Ridge regression, in
  ‘encyclopedia of statistical sciences’, vol. 8},}\ } (\bibinfo {year}
  {1988})\BibitemShut {NoStop}%
\bibitem [{\citenamefont {Zou}\ and\ \citenamefont
  {Hastie}(2005)}]{elasticnet}%
  \BibitemOpen
  \bibfield  {author} {\bibinfo {author} {\bibfnamefont {H.}~\bibnamefont
  {Zou}}\ and\ \bibinfo {author} {\bibfnamefont {T.}~\bibnamefont {Hastie}},\
  }\href {\doibase 10.1111/j.1467-9868.2005.00503.x} {\bibfield  {journal}
  {\bibinfo  {journal} {J. R. Stat. Soc. B}\ }\textbf {\bibinfo {volume}
  {67}},\ \bibinfo {pages} {301} (\bibinfo {year} {2005})}\BibitemShut
  {NoStop}%
\bibitem [{\citenamefont {Papernot}\ \emph {et~al.}(2016)\citenamefont
  {Papernot}, \citenamefont {McDaniel}, \citenamefont {Jha}, \citenamefont
  {Fredrikson}, \citenamefont {Celik},\ and\ \citenamefont
  {Swami}}]{forwardderiv}%
  \BibitemOpen
  \bibfield  {author} {\bibinfo {author} {\bibfnamefont {N.}~\bibnamefont
  {Papernot}}, \bibinfo {author} {\bibfnamefont {P.}~\bibnamefont {McDaniel}},
  \bibinfo {author} {\bibfnamefont {S.}~\bibnamefont {Jha}}, \bibinfo {author}
  {\bibfnamefont {M.}~\bibnamefont {Fredrikson}}, \bibinfo {author}
  {\bibfnamefont {Z.~B.}\ \bibnamefont {Celik}}, \ and\ \bibinfo {author}
  {\bibfnamefont {A.}~\bibnamefont {Swami}},\ }in\ \href {\doibase
  10.1109/EuroSP.2016.36} {\emph {\bibinfo {booktitle} {2016 IEEE European
  Symposium on Security and Privacy (EuroS P)}}}\ (\bibinfo {year} {2016})\
  pp.\ \bibinfo {pages} {372--387}\BibitemShut {NoStop}%
\bibitem [{\citenamefont {Larsen}\ \emph {et~al.}(2017)\citenamefont {Larsen},
  \citenamefont {Mortensen}, \citenamefont {Blomqvist}, \citenamefont
  {Castelli}, \citenamefont {Christensen}, \citenamefont {Du{\l}ak},
  \citenamefont {Friis}, \citenamefont {Groves}, \citenamefont {Hammer},
  \citenamefont {Hargus}, \citenamefont {Hermes}, \citenamefont {Jennings},
  \citenamefont {Jensen}, \citenamefont {Kermode}, \citenamefont {Kitchin},
  \citenamefont {Kolsbjerg}, \citenamefont {Kubal}, \citenamefont {Kaasbjerg},
  \citenamefont {Lysgaard}, \citenamefont {Maronsson}, \citenamefont {Maxson},
  \citenamefont {Olsen}, \citenamefont {Pastewka}, \citenamefont {Peterson},
  \citenamefont {Rostgaard}, \citenamefont {Schi{\o}tz}, \citenamefont
  {Schütt}, \citenamefont {Strange}, \citenamefont {Thygesen}, \citenamefont
  {Vegge}, \citenamefont {Vilhelmsen}, \citenamefont {Walter}, \citenamefont
  {Zeng},\ and\ \citenamefont {Jacobsen}}]{ase}%
  \BibitemOpen
  \bibfield  {author} {\bibinfo {author} {\bibfnamefont {A.~H.}\ \bibnamefont
  {Larsen}}, \bibinfo {author} {\bibfnamefont {J.~J.}\ \bibnamefont
  {Mortensen}}, \bibinfo {author} {\bibfnamefont {J.}~\bibnamefont
  {Blomqvist}}, \bibinfo {author} {\bibfnamefont {I.~E.}\ \bibnamefont
  {Castelli}}, \bibinfo {author} {\bibfnamefont {R.}~\bibnamefont
  {Christensen}}, \bibinfo {author} {\bibfnamefont {M.}~\bibnamefont
  {Du{\l}ak}}, \bibinfo {author} {\bibfnamefont {J.}~\bibnamefont {Friis}},
  \bibinfo {author} {\bibfnamefont {M.~N.}\ \bibnamefont {Groves}}, \bibinfo
  {author} {\bibfnamefont {B.}~\bibnamefont {Hammer}}, \bibinfo {author}
  {\bibfnamefont {C.}~\bibnamefont {Hargus}}, \bibinfo {author} {\bibfnamefont
  {E.~D.}\ \bibnamefont {Hermes}}, \bibinfo {author} {\bibfnamefont {P.~C.}\
  \bibnamefont {Jennings}}, \bibinfo {author} {\bibfnamefont {P.~B.}\
  \bibnamefont {Jensen}}, \bibinfo {author} {\bibfnamefont {J.}~\bibnamefont
  {Kermode}}, \bibinfo {author} {\bibfnamefont {J.~R.}\ \bibnamefont
  {Kitchin}}, \bibinfo {author} {\bibfnamefont {E.~L.}\ \bibnamefont
  {Kolsbjerg}}, \bibinfo {author} {\bibfnamefont {J.}~\bibnamefont {Kubal}},
  \bibinfo {author} {\bibfnamefont {K.}~\bibnamefont {Kaasbjerg}}, \bibinfo
  {author} {\bibfnamefont {S.}~\bibnamefont {Lysgaard}}, \bibinfo {author}
  {\bibfnamefont {J.~B.}\ \bibnamefont {Maronsson}}, \bibinfo {author}
  {\bibfnamefont {T.}~\bibnamefont {Maxson}}, \bibinfo {author} {\bibfnamefont
  {T.}~\bibnamefont {Olsen}}, \bibinfo {author} {\bibfnamefont
  {L.}~\bibnamefont {Pastewka}}, \bibinfo {author} {\bibfnamefont
  {A.}~\bibnamefont {Peterson}}, \bibinfo {author} {\bibfnamefont
  {C.}~\bibnamefont {Rostgaard}}, \bibinfo {author} {\bibfnamefont
  {J.}~\bibnamefont {Schi{\o}tz}}, \bibinfo {author} {\bibfnamefont
  {O.}~\bibnamefont {Schütt}}, \bibinfo {author} {\bibfnamefont
  {M.}~\bibnamefont {Strange}}, \bibinfo {author} {\bibfnamefont {K.~S.}\
  \bibnamefont {Thygesen}}, \bibinfo {author} {\bibfnamefont {T.}~\bibnamefont
  {Vegge}}, \bibinfo {author} {\bibfnamefont {L.}~\bibnamefont {Vilhelmsen}},
  \bibinfo {author} {\bibfnamefont {M.}~\bibnamefont {Walter}}, \bibinfo
  {author} {\bibfnamefont {Z.}~\bibnamefont {Zeng}}, \ and\ \bibinfo {author}
  {\bibfnamefont {K.~W.}\ \bibnamefont {Jacobsen}},\ }\href {\doibase
  10.1088/1361-648x/aa680e} {\bibfield  {journal} {\bibinfo  {journal} {J.
  Phys.: Condens. Matter}\ }\textbf {\bibinfo {volume} {29}},\ \bibinfo {pages}
  {273002} (\bibinfo {year} {2017})}\BibitemShut {NoStop}%
\bibitem [{\citenamefont {Foulkes}\ and\ \citenamefont
  {Haydock}(1989)}]{secondordercorr}%
  \BibitemOpen
  \bibfield  {author} {\bibinfo {author} {\bibfnamefont {W.~M.~C.}\
  \bibnamefont {Foulkes}}\ and\ \bibinfo {author} {\bibfnamefont
  {R.}~\bibnamefont {Haydock}},\ }\href {\doibase 10.1103/PhysRevB.39.12520}
  {\bibfield  {journal} {\bibinfo  {journal} {Phys. Rev. B}\ }\textbf {\bibinfo
  {volume} {39}},\ \bibinfo {pages} {12520} (\bibinfo {year}
  {1989})}\BibitemShut {NoStop}%
\bibitem [{\citenamefont {Elstner}\ \emph {et~al.}(1998)\citenamefont
  {Elstner}, \citenamefont {Porezag}, \citenamefont {Jungnickel}, \citenamefont
  {Elsner}, \citenamefont {Haugk}, \citenamefont {Frauenheim}, \citenamefont
  {Suhai},\ and\ \citenamefont {Seifert}}]{sccextension}%
  \BibitemOpen
  \bibfield  {author} {\bibinfo {author} {\bibfnamefont {M.}~\bibnamefont
  {Elstner}}, \bibinfo {author} {\bibfnamefont {D.}~\bibnamefont {Porezag}},
  \bibinfo {author} {\bibfnamefont {G.}~\bibnamefont {Jungnickel}}, \bibinfo
  {author} {\bibfnamefont {J.}~\bibnamefont {Elsner}}, \bibinfo {author}
  {\bibfnamefont {M.}~\bibnamefont {Haugk}}, \bibinfo {author} {\bibfnamefont
  {T.}~\bibnamefont {Frauenheim}}, \bibinfo {author} {\bibfnamefont
  {S.}~\bibnamefont {Suhai}}, \ and\ \bibinfo {author} {\bibfnamefont
  {G.}~\bibnamefont {Seifert}},\ }\href {\doibase 10.1103/PhysRevB.58.7260}
  {\bibfield  {journal} {\bibinfo  {journal} {Phys. Rev. B}\ }\textbf {\bibinfo
  {volume} {58}},\ \bibinfo {pages} {7260} (\bibinfo {year}
  {1998})}\BibitemShut {NoStop}%
\bibitem [{\citenamefont {Koskinen}\ and\ \citenamefont
  {Mäkinen}(2009)}]{dftbbeginner}%
  \BibitemOpen
  \bibfield  {author} {\bibinfo {author} {\bibfnamefont {P.}~\bibnamefont
  {Koskinen}}\ and\ \bibinfo {author} {\bibfnamefont {V.}~\bibnamefont
  {Mäkinen}},\ }\href {\doibase 10.1016/j.commatsci.2009.07.013} {\bibfield
  {journal} {\bibinfo  {journal} {Comput. Mater. Sci.}\ }\textbf {\bibinfo
  {volume} {47}},\ \bibinfo {pages} {237} (\bibinfo {year} {2009})}\BibitemShut
  {NoStop}%
\bibitem [{\citenamefont {Grundkötter-Stock}\ \emph
  {et~al.}(2012)\citenamefont {Grundkötter-Stock}, \citenamefont {Bezugly},
  \citenamefont {Kunstmann}, \citenamefont {Cuniberti}, \citenamefont
  {Frauenheim},\ and\ \citenamefont {Niehaus}}]{borgset}%
  \BibitemOpen
  \bibfield  {author} {\bibinfo {author} {\bibfnamefont {B.}~\bibnamefont
  {Grundkötter-Stock}}, \bibinfo {author} {\bibfnamefont {V.}~\bibnamefont
  {Bezugly}}, \bibinfo {author} {\bibfnamefont {J.}~\bibnamefont {Kunstmann}},
  \bibinfo {author} {\bibfnamefont {G.}~\bibnamefont {Cuniberti}}, \bibinfo
  {author} {\bibfnamefont {T.}~\bibnamefont {Frauenheim}}, \ and\ \bibinfo
  {author} {\bibfnamefont {T.~A.}\ \bibnamefont {Niehaus}},\ }\href {\doibase
  10.1021/ct200722n} {\bibfield  {journal} {\bibinfo  {journal} {J. Chem.
  Theory Comput.}\ }\textbf {\bibinfo {volume} {8}},\ \bibinfo {pages} {1153}
  (\bibinfo {year} {2012})},\ \bibinfo {note} {pMID: 26593373}\BibitemShut
  {NoStop}%
\bibitem [{\citenamefont {Lian}, \citenamefont {Yoon},\ and\ \citenamefont
  {Lim}(2019)}]{boronfail}%
  \BibitemOpen
  \bibfield  {author} {\bibinfo {author} {\bibfnamefont {M.~H.}\ \bibnamefont
  {Lian}}, \bibinfo {author} {\bibfnamefont {T.~L.}\ \bibnamefont {Yoon}}, \
  and\ \bibinfo {author} {\bibfnamefont {T.~L.}\ \bibnamefont {Lim}},\ }\href
  {\doibase 10.1016/j.cplett.2018.12.023} {\bibfield  {journal} {\bibinfo
  {journal} {Chem. Phys. Lett.}\ }\textbf {\bibinfo {volume} {716}},\ \bibinfo
  {pages} {207} (\bibinfo {year} {2019})}\BibitemShut {NoStop}%
\bibitem [{\citenamefont {Hellström}\ \emph {et~al.}(2013)\citenamefont
  {Hellström}, \citenamefont {Jorner}, \citenamefont {Bryngelsson},
  \citenamefont {Huber}, \citenamefont {Kullgren}, \citenamefont {Frauenheim},\
  and\ \citenamefont {Broqvist}}]{znofail}%
  \BibitemOpen
  \bibfield  {author} {\bibinfo {author} {\bibfnamefont {M.}~\bibnamefont
  {Hellström}}, \bibinfo {author} {\bibfnamefont {K.}~\bibnamefont {Jorner}},
  \bibinfo {author} {\bibfnamefont {M.}~\bibnamefont {Bryngelsson}}, \bibinfo
  {author} {\bibfnamefont {S.~E.}\ \bibnamefont {Huber}}, \bibinfo {author}
  {\bibfnamefont {J.}~\bibnamefont {Kullgren}}, \bibinfo {author}
  {\bibfnamefont {T.}~\bibnamefont {Frauenheim}}, \ and\ \bibinfo {author}
  {\bibfnamefont {P.}~\bibnamefont {Broqvist}},\ }\href {\doibase
  10.1021/jp404095x} {\bibfield  {journal} {\bibinfo  {journal} {The Journal of
  Physical Chemistry C}\ }\textbf {\bibinfo {volume} {117}},\ \bibinfo {pages}
  {17004} (\bibinfo {year} {2013})},\ \bibinfo {note} {pMID:
  23991228}\BibitemShut {NoStop}%
\bibitem [{\citenamefont {Moreira}\ \emph {et~al.}(2009)\citenamefont
  {Moreira}, \citenamefont {Dolgonos}, \citenamefont {Aradi}, \citenamefont
  {da~Rosa},\ and\ \citenamefont {Frauenheim}}]{znorgset}%
  \BibitemOpen
  \bibfield  {author} {\bibinfo {author} {\bibfnamefont {N.~H.}\ \bibnamefont
  {Moreira}}, \bibinfo {author} {\bibfnamefont {G.}~\bibnamefont {Dolgonos}},
  \bibinfo {author} {\bibfnamefont {B.}~\bibnamefont {Aradi}}, \bibinfo
  {author} {\bibfnamefont {A.~L.}\ \bibnamefont {da~Rosa}}, \ and\ \bibinfo
  {author} {\bibfnamefont {T.}~\bibnamefont {Frauenheim}},\ }\href {\doibase
  10.1021/ct800455a} {\bibfield  {journal} {\bibinfo  {journal} {J. Chem.
  Theory Comput.}\ }\textbf {\bibinfo {volume} {5}},\ \bibinfo {pages} {605}
  (\bibinfo {year} {2009})},\ \bibinfo {note} {pMID: 26610226}\BibitemShut
  {NoStop}%
\bibitem [{\citenamefont {Chou}\ \emph {et~al.}(2016)\citenamefont {Chou},
  \citenamefont {Nishimura}, \citenamefont {Fan}, \citenamefont {Mazur},
  \citenamefont {Irle},\ and\ \citenamefont {Witek}}]{siliconfail}%
  \BibitemOpen
  \bibfield  {author} {\bibinfo {author} {\bibfnamefont {C.-P.}\ \bibnamefont
  {Chou}}, \bibinfo {author} {\bibfnamefont {Y.}~\bibnamefont {Nishimura}},
  \bibinfo {author} {\bibfnamefont {C.-C.}\ \bibnamefont {Fan}}, \bibinfo
  {author} {\bibfnamefont {G.}~\bibnamefont {Mazur}}, \bibinfo {author}
  {\bibfnamefont {S.}~\bibnamefont {Irle}}, \ and\ \bibinfo {author}
  {\bibfnamefont {H.~A.}\ \bibnamefont {Witek}},\ }\href {\doibase
  10.1021/acs.jctc.5b00673} {\bibfield  {journal} {\bibinfo  {journal} {J.
  Chem. Theory Comput.}\ }\textbf {\bibinfo {volume} {12}},\ \bibinfo {pages}
  {53} (\bibinfo {year} {2016})},\ \bibinfo {note} {pMID: 26587758}\BibitemShut
  {NoStop}%
\bibitem [{\citenamefont {Sieck}(2000)}]{pbc}%
  \BibitemOpen
  \bibfield  {author} {\bibinfo {author} {\bibfnamefont {A.}~\bibnamefont
  {Sieck}},\ }\emph {\bibinfo {title} {{Structure and physical properties of
  silicon clusters and of vacancy clusters in bulk silicon}}},\ \href
  {https://nbn-resolving.de/urn:nbn:de:hbz:466-20000101252} {\bibinfo {type}
  {Dissertation}},\ \bibinfo  {school} {University of Paderborn} (\bibinfo
  {year} {2000})\BibitemShut {NoStop}%
\bibitem [{\citenamefont {{Markov}}\ \emph {et~al.}(2015)\citenamefont
  {{Markov}}, \citenamefont {{Aradi}}, \citenamefont {{Yam}}, \citenamefont
  {{Xie}}, \citenamefont {{Frauenheim}},\ and\ \citenamefont
  {{Chen}}}]{siband1}%
  \BibitemOpen
  \bibfield  {author} {\bibinfo {author} {\bibfnamefont {S.}~\bibnamefont
  {{Markov}}}, \bibinfo {author} {\bibfnamefont {B.}~\bibnamefont {{Aradi}}},
  \bibinfo {author} {\bibfnamefont {C.}~\bibnamefont {{Yam}}}, \bibinfo
  {author} {\bibfnamefont {H.}~\bibnamefont {{Xie}}}, \bibinfo {author}
  {\bibfnamefont {T.}~\bibnamefont {{Frauenheim}}}, \ and\ \bibinfo {author}
  {\bibfnamefont {G.}~\bibnamefont {{Chen}}},\ }\href {\doibase
  10.1109/TED.2014.2387288} {\bibfield  {journal} {\bibinfo  {journal} {IEEE
  Transactions Electron Devices}\ }\textbf {\bibinfo {volume} {62}},\ \bibinfo
  {pages} {696} (\bibinfo {year} {2015})}\BibitemShut {NoStop}%
\bibitem [{\citenamefont {Kresse}\ and\ \citenamefont
  {Furthm\"uller}(1996)}]{VASP4}%
  \BibitemOpen
  \bibfield  {author} {\bibinfo {author} {\bibfnamefont {G.}~\bibnamefont
  {Kresse}}\ and\ \bibinfo {author} {\bibfnamefont {J.}~\bibnamefont
  {Furthm\"uller}},\ }\href {\doibase 10.1103/PhysRevB.54.11169} {\bibfield
  {journal} {\bibinfo  {journal} {Phys. Rev. B}\ }\textbf {\bibinfo {volume}
  {54}},\ \bibinfo {pages} {11169} (\bibinfo {year} {1996})}\BibitemShut
  {NoStop}%
\bibitem [{\citenamefont {Monkhorst}\ and\ \citenamefont
  {Pack}(1976)}]{MonkhorstPack}%
  \BibitemOpen
  \bibfield  {author} {\bibinfo {author} {\bibfnamefont {H.~J.}\ \bibnamefont
  {Monkhorst}}\ and\ \bibinfo {author} {\bibfnamefont {J.~D.}\ \bibnamefont
  {Pack}},\ }\href {\doibase 10.1103/PhysRevB.13.5188} {\bibfield  {journal}
  {\bibinfo  {journal} {Phys. Rev. B}\ }\textbf {\bibinfo {volume} {13}},\
  \bibinfo {pages} {5188} (\bibinfo {year} {1976})}\BibitemShut {NoStop}%
\bibitem [{\citenamefont {Kresse}\ and\ \citenamefont {Hafner}(1993)}]{VASP1}%
  \BibitemOpen
  \bibfield  {author} {\bibinfo {author} {\bibfnamefont {G.}~\bibnamefont
  {Kresse}}\ and\ \bibinfo {author} {\bibfnamefont {J.}~\bibnamefont
  {Hafner}},\ }\href {\doibase 10.1103/PhysRevB.47.558} {\bibfield  {journal}
  {\bibinfo  {journal} {Phys. Rev. B}\ }\textbf {\bibinfo {volume} {47}},\
  \bibinfo {pages} {558} (\bibinfo {year} {1993})}\BibitemShut {NoStop}%
\bibitem [{\citenamefont {Kresse}\ and\ \citenamefont {Hafner}(1994)}]{VASP2}%
  \BibitemOpen
  \bibfield  {author} {\bibinfo {author} {\bibfnamefont {G.}~\bibnamefont
  {Kresse}}\ and\ \bibinfo {author} {\bibfnamefont {J.}~\bibnamefont
  {Hafner}},\ }\href {\doibase 10.1103/PhysRevB.49.14251} {\bibfield  {journal}
  {\bibinfo  {journal} {Phys. Rev. B}\ }\textbf {\bibinfo {volume} {49}},\
  \bibinfo {pages} {14251} (\bibinfo {year} {1994})}\BibitemShut {NoStop}%
\bibitem [{\citenamefont {Kresse}\ and\ \citenamefont
  {Furthmüller}(1996)}]{VASP3}%
  \BibitemOpen
  \bibfield  {author} {\bibinfo {author} {\bibfnamefont {G.}~\bibnamefont
  {Kresse}}\ and\ \bibinfo {author} {\bibfnamefont {J.}~\bibnamefont
  {Furthmüller}},\ }\href {\doibase 10.1016/0927-0256(96)00008-0} {\bibfield
  {journal} {\bibinfo  {journal} {Comput. Mater. Sci.}\ }\textbf {\bibinfo
  {volume} {6}},\ \bibinfo {pages} {15} (\bibinfo {year} {1996})}\BibitemShut
  {NoStop}%
\end{thebibliography}%

\end{document}